\documentclass[prd,floatfix,onecolumn,amsmath,amssymb,10pt]{revtex4}
\usepackage{newtxtext,newtxmath} 
\usepackage{graphicx,color,dcolumn,booktabs,bm}
\usepackage{subfigure}
\usepackage{amssymb}
\usepackage{longtable}
\usepackage{indentfirst}
\usepackage{epsfig,gensymb,siunitx}
\usepackage{feynmf}   %{feynmp}
\usepackage{epstopdf}   %{feynmp}
\usepackage{slashed}  %for Feynman symbols
\usepackage{cases}
\usepackage{xcolor}
\definecolor{navyblue}{RGB}{0,0,150}
\definecolor{CobaltBlue}{rgb}{0,0.28,.67}
\definecolor{maroon}{RGB}{139,25,150}%burada 0-255 arasi her biri icin numara vererek renk elde et
\usepackage{multirow}
\usepackage{float}
\usepackage{pgf}
\usepackage{graphicx,color,dcolumn,booktabs,bm}
\usepackage[colorlinks, citecolor=purple,anchorcolor=purple,menucolor=purple, linkcolor=purple,filecolor=purple,runcolor=purple,urlcolor=purple,frenchlinks=purple, urlcolor=purple]{hyperref}
\usepackage{orcidlink}
\usepackage{bbold}
\usepackage{tikz}
\usepackage[compat=1.1.0]{tikz-feynman}
\usetikzlibrary{decorations.markings}

\tikzfeynmanset{
	momentum arrow/.style={
		decoration={
			markings,
			mark=at position 0.55 with {\arrowreversed{>}}
		},
		postaction={decorate}
	}
}

\begin{document}
		\preprint{}
	
	\title{\color{navyblue}{Photon and gluon gravitational form factors of proton in QCD}}
	
	\author{Z.~Asmaee$^{1}$\orcidlink{0000-0002-3357-0574}}
	\email{zahra.asmaee@ut.ac.ir}
	\thanks{Corresponding author}
	
	\author{K.~Azizi$^{1,2}$\orcidlink{0000-0003-3741-2167}}
	\email{kazem.azizi@ut.ac.ir}
	\thanks{Corresponding author}
	
	\affiliation{
		$^{1}$Department of Physics, \href{https://ut.ac.ir/en}{University of Tehran}, North Karegar Avenue, Tehran 14395-547, Iran\\
		$^{2}$Department of Physics, \href{https://www.dogus.edu.tr/en}{Dogus University}, Dudullu-\"{U}mraniye, 34775
		Istanbul,  T\"{u}rkiye}	
%	\date{\today}
\begin{abstract}
The energy momentum tensor allows us to probe the internal gravitational structure of the proton. The corresponding gravitational form factors contain information about its different contributions.
In this work, we investigate the photon and gluon gravitational form factors of the proton within the QCD sum rule approach. Four independent gravitational form factors are computed in the range $0<Q^2<2\ \mathrm{GeV}^2$. The mass and scalar radii of the photon and gluon sectors are obtained in the forward limit. We find that the gluon mass radius is smaller than the photon mass radius.
These results may be  useful  for future lattice QCD calculations and phenomenological studies.
\end{abstract}
	
	\maketitle
	
	%%%%%%%%%%%%%%%%%%%%%%%%%%%%%%%%%%%%%%%%%%%%%%%%%%%%%%%%%%%%%%%%%%%%%%%%%%%%%%%%%%%
	\section{Introduction}\label{Introduction} 
The pioneering studies of the proton structure through electron scattering experiments~\cite{Hofstadter:1953zjy, Hofstadter:1956qs} and deep inelastic scattering measurements at the Stanford Linear Accelerator Center~\cite{Bloom:1969kc, Breidenbach:1969kd}, together with the development of quantum chromodynamics (QCD)~\cite{Fritzsch:1973pi, Gross:1973id, Politzer:1973fx, Gell-Mann:1964ewy, Zweig:1964jf}, established the modern framework for understanding hadrons in terms of their quark and gluon degrees of freedom. Nevertheless, a complete description of hadron structure requires not only an understanding of partonic dynamics but also a characterization of how energy, momentum, and internal forces are distributed within hadrons. This information is encoded in the gravitational form factors (GFFs), which are defined through the matrix elements of the QCD energy momentum tensor (EMT) and provide access to the spatial distributions of energy, momentum, angular momentum, and mechanical properties inside hadrons. Their precise determination therefore plays a crucial role in revealing the underlying structure of hadrons.

Among hadrons, the proton represents a particularly important case, as the lightest stable baryon, and has been the subject of extensive experimental and theoretical studies. In recent years, significant efforts have been devoted to the precise determination of its GFFs, as these quantities provide detailed information on the distributions of energy and momentum as well as the internal mechanical structure of the proton, including the spatial distributions of pressure and shear forces. Moreover, the GFFs encode essential aspects of proton structure, such as the mass decomposition, angular momentum content, the mechanical properties described by the $D$-term, and the corresponding spatial radii. Their precise determination is therefore crucial for obtaining a comprehensive understanding of the proton's internal dynamics.

Beyond their role in characterizing the internal structure of the proton, GFFs are also connected to experimentally accessible partonic observables through generalized parton distributions (GPDs). The GFFs associated with the symmetric traceless part of the EMT are related to the second Mellin moments of GPDs~\cite{Ji:1996nm, Muller:1994ses, Radyushkin:1996nd}. This relation establishes a connection between GFFs and hard exclusive processes, including deeply virtual Compton scattering (DVCS)~\cite{Ji:1996ek, dHose:2016mda, Kumericki:2016ehc} and exclusive meson production~\cite{Collins:1996fb, Mankiewicz:1997bk}. In particular, DVCS measurements at the Thomas Jefferson National Accelerator Facility (JLab) have provided important constraints on the quark contributions to the gravitational structure of the proton~\cite{Burkert:2018bqq, Burkert:2021ith}. These developments have established hard exclusive processes as a powerful tool for exploring the energy, momentum, and mechanical properties of hadrons through their underlying partonic degrees of freedom.

Complementary to these experimental efforts, theoretical approaches have played a central role in exploring the GFFs of the proton. Lattice QCD has provided valuable information on the EMT matrix elements~\cite{Gockeler:2003jfa, Hagler:2003jd, Shanahan:2018nnv}, while other nonperturbative approaches, including chiral effective field theory~\cite{Dorati:2007bk, Chen:2001pva}, light-cone QCD sum rules~\cite{Azizi:2019ytx, Dehghan:2025ncw}, holographic QCD~\cite{Mamo:2022eui, Fujita:2022jus}, and the Skyrme model~\cite{Cebulla:2007ei, Kim:2012ts}, have provided complementary predictions for the gravitational properties of the proton. These theoretical studies have significantly advanced our knowledge of the proton's gravitational structure and provided valuable predictions for its nonperturbative properties, establishing GFFs as key observables for exploring QCD dynamics beyond the perturbative regime.

A further important aspect of the gravitational structure of hadrons is the decomposition of the QCD energy momentum tensor (EMT) into its individual dynamical sectors. The QCD EMT can be separated into quark and gluon sectors, whose combined contributions determine the gravitational structure of hadrons. Among these sectors, the gluon contribution is of particular importance due to its essential role in the nonperturbative dynamics of QCD. Nevertheless, accessing the gluonic structure of hadrons remains particularly challenging, since gluons cannot be directly probed through electromagnetic interactions. Consequently, gluon gravitational form factors (GFFs) provide an essential tool for investigating the role of gluons in the energy momentum structure of hadrons.

Among these sectors, the gluon contribution remains one of the least explored components of the proton gravitational structure. Although substantial progress has been achieved in determining the quark and total proton GFFs, considerably less is known about the gluon sector. Gluon GFFs have been investigated within several theoretical approaches, including lattice QCD calculations~\cite{Alexandrou:2018xnp, Alexandrou:2020sml, Shanahan:2018pib, Yang:2018bft, Yang:2018nqn, Alexandrou:2017oeh, Pefkou:2021fni, Hackett:2023rif} and holographic QCD models~\cite{Mamo:2022eui, Duran:2022xag}, providing valuable insight into the gluonic contribution to the energy momentum structure and gravitational properties of the proton~\cite{Tong:2021ctu, Tandy:2025tea}. From the experimental perspective, direct constraints on gluon GFFs remain challenging due to the absence of direct electromagnetic probes of gluons. Nevertheless, exclusive heavy quarkonium production processes, such as $J/\psi$ and $\Upsilon$ photo- and leptoproduction~\cite{Mamo:2019mka, Hatta:2018ina, Boussarie:2020vmu, Wang:2022ndz}, have been proposed as promising probes of the gluonic gravitational structure of the proton. These processes are expected to provide important constraints on gluon GFFs through ongoing experiments at JLab~\cite{GlueX:2019mkq, Guo:2023pqw} and future measurements at the Electron-Ion Collider (EIC)~\cite{AbdulKhalek:2021gbh}.

While the gluon sector has been extensively investigated, the photon contribution to the gravitational structure of the proton remains much less explored. The photon component of the EMT provides an additional contribution to the GFFs and reflects the role of electromagnetic degrees of freedom in the energy momentum structure of the proton. Recently, the photon contribution to the momentum GFF has been investigated within the Weizs\"acker--Williams framework, providing new insight into the electromagnetic component of the gravitational structure of the proton~\cite{Hagiwara:2024wqz}. Moreover, the separate photon contribution to GFFs has been studied in quantum electrodynamics, where the photon and fermion contributions to the EMT can be consistently separated~\cite{Freese:2022jlu}. However, a systematic determination of the complete photon GFFs of composite hadrons remains largely unexplored. Investigating this contribution is essential for achieving a more complete understanding of the electromagnetic role in the gravitational structure of the proton and the interplay between strong and electromagnetic dynamics.

Despite recent progress in understanding the individual contributions to the proton gravitational structure, a unified description of the different dynamical sectors of the proton EMT remains incomplete. In particular, the simultaneous investigation of the gluon and photon contributions to the proton GFFs within a unified nonperturbative framework remains an open problem. Such a study is essential for achieving a more complete understanding of how different fundamental degrees of freedom contribute to the energy momentum structure of the proton.

In the present work, we investigate the photon and gluon contributions to the proton GFFs within the framework of QCD sum rules (QCDSR). In Sec.~\ref{sec:EMT}, we introduce the gluon and photon components of the EMT and present the corresponding parametrization of the proton matrix element in terms of the GFFs. The derivation of the QCDSR is presented in Sec.~\ref{sec:formalism}, where the three-point correlation function is analyzed in both the hadronic and QCD representations. The numerical analysis, including the determination of the GFFs, their momentum-transfer dependence, and the extraction of the mass and scalar radii for the photon and gluon sectors, is presented in Sec.~\ref{sec:numerical}. Finally, our conclusions are summarized in Sec.~\ref{sec:conclusion}.

\section{EMT and Proton Matrix Element}\label{sec:EMT}
The proton GFFs are obtained from the matrix elements of the EMT. In this work, we study the photon and gluon contributions to the gravitational form factors of the proton. We first introduce the corresponding components of the EMT and then present the Lorentz decomposition of their matrix elements between proton states.
The gluon part of the EMT is given by~\cite{Dehghan:2025ncw,Burkert:2023wzr},
\allowdisplaybreaks
\begin{align}
T_{\mu\nu}^g(x)=-G^a_{\mu\rho}(x)G^{a\hspace{0.8mm} \rho}_{\nu}(x)+\frac{1}{4}g_{\mu\nu}G^a_{\rho\delta}(x)
G^{a \, \rho\delta}(x),
\label{EMTcurrentgluon}
\end{align}
where $g_{\mu\nu}=\mathrm{diag}(+1,-1,-1,-1)$ is the Minkowski metric, and the gluon field strength tensor $G_{\mu\nu}^{a}(x)$ is defined as
\allowdisplaybreaks
\begin{align}
G_{\mu\nu}^{a}(x)=\partial_{\mu}A_{\nu}^{a}(x)-\partial_{\nu}A_{\mu}^{a}(x)+g_s f^{abc}A_{\mu}^{b}(x)A_{\nu}^{c}(x),
\label{field strength tensor}
\end{align}
here, $A_\mu^a$ and $g_s$ denote the gluon gauge field and the strong coupling constant, respectively.
The structure constants of the SU(3)$_c$ color group are given by $f^{abc}$, where $a$, $b$, and $c$ are color indices.
Similarly, the gauge invariant electromagnetic contribution to the EMT is~\cite{Hagiwara:2024wqz, Freese:2022jlu},
\allowdisplaybreaks
\begin{align}
T_{\mu\nu}^\gamma(x)=-F_{\mu\rho}(x)F^{\hspace{0.9mm}\rho}_{\nu}(x)+\frac{1}{4}g_{\mu\nu}F_{\rho\delta}(x)F^{\rho\delta}(x).
\label{EMTcurrentphoton}
\end{align}
Unlike the gluon field strength tensor, the electromagnetic field strength tensor $F_{\mu\nu}$ has no color index or self interaction terms, and is written as
\allowdisplaybreaks
\begin{align}
F_{\mu\nu}(x)=\partial_{\mu}A_{\nu}(x)-\partial_{\nu}A_{\mu}(x),
\label{field strength tensor-photon}
\end{align}
where $A_\mu$ denotes the electromagnetic gauge field.

The proton matrix element of each EMT component can be parametrized in terms of four independent GFFs as~\cite{Azizi:2019ytx, Dehghan:2025ncw, Burkert:2023wzr, Kopeliovich:2010xm,  Ji:1996ek, Bakker:2004ib, Polyakov:2018zvc}
\allowdisplaybreaks
\begin{align}
\langle N(p',s')|T_{\mu\nu}^j(x)|N(p,s)\rangle=\bar{u}(p',s')&\Bigg\{\frac{P_\mu P_\nu}{m} A^j(t)
+\frac{i}{2}\ \frac{\left(P_\mu \sigma_{\nu\rho}+P_\nu \sigma_{\mu\rho}\right)\Delta^\rho}{m} J^j(t)
\nonumber\\&
+\frac{\left(\Delta_\mu \Delta_\nu-g_{\mu\nu}\Delta^2\right)}{4m}D^j(t)+m g_{\mu\nu}\ \bar{c}^j(t)\Bigg\}u(p,s)\ e^{i(p'-p)\cdot x},
\label{matrix element}
\end{align}	
where $j$ denotes the corresponding contribution to the EMT, with $j=g,\gamma$ representing the gluon and photon sectors, respectively. 
The spinors $u(p,s)$ and $\bar{u}(p',s')$ describe the initial and final proton states with four momenta  $p$ and $p'$, and spins $s$ and $s'$, respectively. 
The total momentum and momentum transfer are defined as $P=(p+p')/2$ and $\Delta=p'-p$, with $t=\Delta^2$. Here, $\sigma_{\mu\rho}=\frac{i}{2}[\gamma_\mu,\gamma_\rho]$, and $m$ denotes the proton mass. 
The functions  $A^{j}(t)$, $J^{j}(t)$, $D^{j}(t)$, and $\bar{c}^{j}(t)$ are the GFFs for each EMT component.
The form factor $\bar{c}^{j}(t)$ describes the non-conservation contributions.
These form factors (FFs) provide information about the gravitational structure of the proton and its spatial properties, such as the mass and scalar radii.

\section{Photon and Gluon GFFs of the Proton in QCDSR}\label{sec:formalism} 
To investigate the photon and gluon GFFs of the proton within the QCDSR framework, a three-point correlation function involving the proton interpolating currents and the corresponding photon and gluon EMT operators is written as
\allowdisplaybreaks
\begin{align}
\Pi^j_{\mu\nu}(p,p') = i^2 \int d^4 x\ e^{-ip\cdot x}\int d^4 y\ e^{ip'\cdot y}\langle 0 |\mathcal{T}[J_N(y)T^{j}_{\mu \nu}(0)\bar{J}_N(x)]| 0 \rangle,
\label{corrf}
\end{align}
where $\mathcal{T}$ denotes the time ordering operator, and the general form of the proton interpolating current $J_N$ is \cite{Dehghan:2025ncw, Khodjamirian:2020btr, Colangelo:2000dp},
\allowdisplaybreaks
\begin{align}
J_{N}(x)=2 \varepsilon_{abc}\Big[\big(u^{aT}(x) C d^{b}(x)\big)\gamma_{5}u^{c}(x)+\beta\big(u^{aT}(x) C \gamma_{5} d^{b}(x)\big)u^{c}(x)\Big],
\label{interpolating}
\end{align}
where $C$ is the charge conjugation matrix, $u(x)$ and $d(x)$ are the up- and down-quark fields, respectively. $\varepsilon_{abc}$ is the totally antisymmetric color tensor, and  $\beta$ is an arbitrary mixing parameter.

In the QCDSR approach, the correlation function in Eq.~\eqref{corrf} is calculated in two different descriptions. 
On the hadronic (physical) side, the correlation function is written in terms of proton states and their GFFs, while on the QCD side, it is calculated using quark and gluon degrees of freedom.
To suppress the contributions of excited states and the continuum and isolate the ground state proton contribution, double Borel transformations with respect to $p^2$ and $p'^2$ are applied to both sides, and then continuum subtraction is performed.
The QCDSRs for the GFFs are obtained by matching the coefficients of the same Lorentz structures on the hadronic and QCD sides.
\subsection{Physical Representation}\label{physical side}
To obtain the hadronic representation of the correlation function, a complete set of intermediate hadronic states with the same quantum numbers as the proton interpolating current is inserted into Eq.~\eqref{corrf} between the interpolating currents and the EMT operator~\cite{Dehghan:2025ncw, Dehghan:2023ytx, Dehghan:2025eov, Asmaee:2025elo, Asmaee:2026wrk},
\allowdisplaybreaks
\begin{align}
1=\vert 0\rangle\langle0\vert +\sum_{h}\int\frac{d^4 p_h}{(2\pi)^4}(2\pi) \delta(p^2_h-m^2)|N(p_h)\rangle  \langle N(p_h)|+\mbox{higher Fock states},
\label{set}
\end{align}
where $|N(p_h)\rangle$ represents an intermediate proton state with four momentum $p_h$. 
The sum runs over all intermediate states, including the ground state proton and excited states. 
After inserting Eq.~\eqref{set} into Eq.~\eqref{corrf} and performing the integrations over the space-time coordinates $x$ and $y$, one obtains the ground state proton contribution using the identity~\cite{Dehghan:2025ncw, Dehghan:2023ytx, Dehghan:2025eov, Asmaee:2025elo, Asmaee:2026wrk},
\allowdisplaybreaks
\begin{align}
\int d^4 x\int\frac{d^4 p_h}{(2\pi)^4}(2\pi) \delta(p^2_h-m^2)e^{i(p_h - p)\cdot x}= \frac{i}{m^2 - p^2}.
\label{identity}
\end{align}
Using Eq.~\eqref{identity} and performing the calculations, the hadronic representation of the correlation function in Eq.~\eqref{corrf} is written as~\cite{ Dehghan:2023ytx, Dehghan:2025eov, Asmaee:2025elo, Asmaee:2026wrk}:
\allowdisplaybreaks
\begin{align}
\Pi_{\mu\nu}^{\mathrm{phy},\ j}(p,p')=\sum_{s}\ \sum_{s'}\frac{\langle0|J_N(0)|N(p',s')\rangle\langle N(p',s')|T_{\mu\nu}^{j}(0)|N(p,s)\rangle\langle N(p,s)|\bar{J}_N(0)|0\rangle}{(m^2-p'^2)(m^2-p^2)}+\cdots,
\label{equation phys}
\end{align}
where the dots represent the contributions from excited states and the continuum.
The first and third matrix elements in Eq.~\eqref{equation phys} are written as~\cite{Aliev:2011ku, Aliev:2016jnp, Ioffe:1981kw, Aliev:2002ra, Azizi:2014yea},
\allowdisplaybreaks
\begin{align}
\langle0|J_N(0)|N(p',s')\rangle=\lambda_N\ u(p',s'),
\label{residue}
\end{align}
where the proton residue $\lambda_N$ describes the coupling strength between the interpolating current and the proton state.
After substituting Eqs.~\eqref{matrix element} and \eqref{residue} into Eq.~\eqref{equation phys} and summing over spins using the completeness relation,
\allowdisplaybreaks
\begin{align}
\sum_{s} u(p,s)\bar{u}(p,s)=\slashed{p}+m,
\label{equation complete}
\end{align}
the physical representation of the correlation function is expressed in terms of the proton GFFs as,
\allowdisplaybreaks
\begin{align}
\Pi_{\mu\nu}^{\mathrm{phy},\ j}(p,p')=\frac{\left| \lambda_N\right| ^2}{(m^2-p'^2)(m^2-p^2)}\left(\slashed{p}' + m\right)&\Bigg\{\frac{P_\mu P_\nu}{m} A^j(t)+\frac{i}{2}\ \frac{\left(P_\mu \sigma_{\nu\rho}+P_\nu \sigma_{\mu\rho}\right)\Delta^\rho}{m} J^j(t)
\nonumber\\&
+\frac{\left(\Delta_\mu \Delta_\nu-g_{\mu\nu}\Delta^2\right)}{4m}D^j(t)+m g_{\mu\nu}\ \bar{c}^j(t)\Bigg\}
\left(\slashed{p}+m\right)+\cdots.
\label{equation hadronic}
\end{align}	
Finally, to suppress the contributions from excited states and the continuum, a double Borel transformation with respect to $p^2$ and $p'^2$ is applied with the corresponding Borel parameters $M_1^2$ and $M_2^2$, respectively
~\cite{Colangelo:2000dp, Ozdem:2017jqh, Azizi:2018duk},
\allowdisplaybreaks
\begin{align}
\mathcal{B}_{M_1^2}	\mathcal{B}_{M_2^2}\Bigg[ \frac{1}{\left( m^2-p^{2}\right) \left( m^2-p^{\prime 2}\right) }\Bigg] =e^{-\frac{m^2}{M_1^2}}e^{-\frac{m ^2}{M_2^2}}=e^{-\frac{m^2}{M^2}}.
\label{Borel}
\end{align}
Since the initial and final states are the same proton, the Borel parameters are equal $M_1^2=M_2^2$. Thus, $M_1^2=M_2^2=2M^2$ is used in the last step of Eq.~\eqref{Borel}.
After applying the Borel transformation in Eq.~\eqref{Borel}, the hadronic representation of the correlation function becomes
\allowdisplaybreaks
\begin{align}
	\Pi_{\mu\nu}^{\mathrm{phy},\ j}(Q^2)=\left| \lambda_N\right| ^2\ e^{-m^2/M^2}&\ \Bigg[
	\Pi_1^{\mathrm{phy}, \ j}(Q^2)\ p_\mu p_\nu \mathbb{1}
	+\Pi_2^{\mathrm{phy}, \ j}(Q^2)\ p_\mu p'_\nu \mathbb{1}
	+\Pi_3^{\mathrm{phy}, \ j}(Q^2)\ p_\mu p_\nu \slashed{p}'
	+\Pi_4^{\mathrm{phy}, \ j}(Q^2)\ p'_\mu p'_\nu \slashed{p}
	\nonumber\\&
	+\Pi_{5}^{\mathrm{phy}, \ j}(Q^2)\ g_{\mu\nu}\slashed{p}
	+\Pi_{6}^{\mathrm{phy}, \ j}(Q^2)\ p_\mu p_\nu \slashed{p}\slashed{p}'
	+\Pi_{7}^{\mathrm{phy}, \ j}(Q^2)\ p_\nu p'_\mu \slashed{p}\slashed{p}'
	+\Pi_{8}^{\mathrm{phy}, \ j}(Q^2)\ p_\mu p'_\nu \slashed{p}\slashed{p}'
	\nonumber\\&
	+\Pi_{9}^{\mathrm{phy}, \ j}(Q^2)\ p'_\mu p'_\nu \slashed{p}\slashed{p}'
	+\Pi_{10}^{\mathrm{phy}, \ j}(Q^2)\ g_{\mu\nu}\slashed{p}\slashed{p}'
	+\dots\Bigg],
	\label{equation correlation physical}
\end{align}
where $Q^2=-t$ and $\mathbb{1}$ denotes the identity matrix. 
The invariant functions $\Pi_{i}^{\mathrm{phy}, \ j}(Q^2)$ (with $i=1, \cdots, 10$) are obtained in terms of the proton GFFs and its physical parameters and their explicit forms are presented in Appendix~\ref{appA}. 
For the numerical analysis, the Lorentz structures used to calculate the photon and gluon GFFs of the proton are retained, while the remaining structures are represented by dots.
The proton residue $\lambda_N$ in the hadronic representation is obtained from the QCDSR analys is~\cite{Aliev:2002ra, Aliev:2011ku},
\begin{align}
\left| \lambda_N\right| ^2\ e^{-m^2/M^2}= \Bigg\{\frac{M^6}{256 \pi^4} E_2(x) (5+2 \beta + \beta^2) 
	- \frac{\langle \bar{q}q \rangle^2}{6} \Big[6 (1-\beta^2)  -
	(1-\beta)^2  \Big] + \frac{m_0^2}{24 M^2} \langle \bar{q}q \rangle^2 \Big[12 (1-\beta^2) - (1-\beta)^2  \Big]\Bigg\}, 
		\label{residue1}
\end{align}
where $\langle\bar{q}q\rangle$ is the quark condensate and $m_0^2$ is the mixed quark-gluon condensate parameter, with
$\langle\bar{q}g_s\sigma G q\rangle = m_0^2 \langle\bar{q}q\rangle$.
The function $E_2(x)$ is the continuum subtraction factor~\cite{Dehghan:2025ncw},
\allowdisplaybreaks
\begin{align}
E_2(x)=1-e^{-x}\left(1+x+\frac{x^2}{2}\right),
\end{align}
here, $x=s_0/M^2$, and $s_0$ is the threshold parameter separating the ground state contribution from the continuum.

\subsection{QCD Representation}\label{QCD side}
In this section, the correlation function is calculated in terms of quark and gauge fields.
First, the EMT operators in Eqs.~\eqref{EMTcurrentgluon} and \eqref{EMTcurrentphoton} and the proton interpolating current in Eq.~\eqref{interpolating} are substituted into the correlation function in Eq.~\eqref{corrf}. Next, Wick's theorem is applied to the time ordered product to obtain the QCD side.
Since the gluon and photon parts of the EMT are different, the QCD side is calculated separately in Subsections~\ref{QCD-gluon} and \ref{QCD-photon}. 

\subsubsection{QCD representation of the gluon contribution}\label{QCD-gluon}
In this subsection, the gluon contribution to the correlation function on the QCD side is investigated. The calculation starts with the vacuum matrix element
$\left\langle 0 \left|\mathcal{T}\left[J_N(y)T_{\mu\nu}^{g}(0)\bar{J}_N(x)\right]\right|0\right\rangle $
in Eq.~\eqref{corrf}.
By substituting the gluon EMT from Eq.~\eqref{EMTcurrentgluon} and the proton interpolating current and its conjugate from Eq.~\eqref{interpolating}, the vacuum matrix element is written in terms of quark and gluon fields. Then, Wick's theorem is applied to obtain different contractions, which are classified into two distinct classes.

\vspace{0.2cm}
\textbf{For the first class} of contributions, the quark fields in the proton interpolating currents are contracted with each other and the result is written in terms of light-quark propagators that depend on $(y-x)$ (see Eq.~\eqref{Pi1}). 
The two gluon field strength tensors in the EMT are contracted with each other. Using Eqs.~\eqref{EMTcurrentgluon} and \eqref{condensation gluon}, the vacuum matrix element of the gluon EMT vanishes. Thus, this class has no contribution.

\begin{figure}[ht]
\centering
\includegraphics[height=3cm, width=6.8cm]{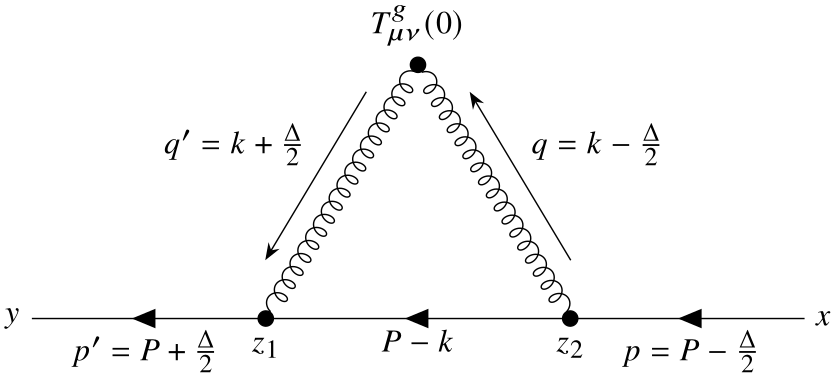}
\caption{Schematic diagram of the gluon EMT insertion at the quark-gluon vertex.}	\label{fig:gluon-correction}
\end{figure}

\vspace{0.2cm}
\textbf{For the second class} of contributions, the gluon fields in the EMT interact with the quark fields in the proton interpolating currents at the quark-gluon vertex. This interaction changes the light-quark propagator due to the gluon EMT insertion. This contribution is shown in Fig.~\ref{fig:gluon-correction}.

In this class of contributions, the gluon EMT insertion modifies the light-quark propagators as
\allowdisplaybreaks
\begin{align}
S^{ab}_\psi\rightarrow  S^{ab}_\psi+\Delta S^{ab,\ g}_\psi,
\label{correction}
\end{align}
where $S^{ab}_\psi$ is the light-quark propagator for the first class, while $\Delta S^{ab,\ g}_\psi$ describes the correction due to the gluon EMT insertion in the second class. As discussed above, the first class has no contribution to the correlation function, and only the second class contributes.
The correction $\Delta S^{ab,\ g}_\psi(x)$ in Eq.~\eqref{correction} is calculated using Green's functions in the interaction picture,
\allowdisplaybreaks
\begin{align}
\left\langle 0 \left|\mathcal{T}\left[q(y)T_{\mu\nu}^{g}(0)\bar q(x)\right]\right|0\right\rangle_{\mathrm{int}}=\frac{\left\langle 0 \left|\mathcal{T}\left[q(y)T_{\mu\nu}^{g}(0)\bar q(x)\exp \left(i\int d^4z\ \mathcal{L}^{qg}_{\mathrm{int}}(z)\right)\right]\right|0\right\rangle}{\left\langle 0 \left|\mathcal{T}\left[\exp \left(i\int d^4z\ \mathcal{L}^{qg}_{\mathrm{int}}(z)\right)\right]\right|0\right\rangle},
\label{Green function}
\end{align}
where $q(x)$ denotes the light-quark field. To calculate the correction, only the term of the gluon EMT that contains two gluon fields is used. The nonlinear terms in the gluon field strength tensor that describe three- and four-gluon interactions are neglected. The quark-gluon interaction is described by the Lagrangian
\allowdisplaybreaks
\begin{align}
\mathcal{L}^{qg}_{\mathrm{int}}(z)=-g_s \bar{q}^i(z)\gamma^\alpha (T^e)_{ij} q^j(z) A^e_\alpha(z).
\label{gluon lag}
\end{align}
When the exponential in Eq.~\eqref{Green function} is expanded in powers of the interaction Lagrangian, the zeroth- and first order terms do not contribute to the correlation function. The first nonzero contribution comes from the second order term,
$i^2\Big(\int d^{4}z\ \mathcal{L}^{qg}_{\mathrm{int}}(z)\Big)^{2}/2!$.
This second-order term contains two $qqg$ vertices. The gluon fields at these vertices are contracted with the gluon fields in the gluon EMT. 
%As a result, the gluonic EMT insertion is effectively included in the correlation function through the correction term $\Delta S_\psi^{ab,g}$ introduced in Eq.~\eqref{correction}. 
The explicit form of this correction to the light-quark propagator is obtained from Eq.~\eqref{Green function}.
As an example, the correction to the up-quark propagator $S_u^{ca'}(y-x)$ is obtained as
\allowdisplaybreaks
\begin{align}
\Delta S_{u}^{ca',\ g}(y-x)=-\frac{g_s^2}{2}\int d^4z_1 d^4z_2 &\Big<0\Big|	\mathcal{T}\Big[u^c(y) T_{\mu\nu}^{g}(0) \bar{u}^{a'}(x) \bar{u}^i(z_1)\gamma^\alpha (T^e)_{ij} u^j(z_1) A^e_\alpha(z_1)
\bar{u}^k(z_2)\gamma^\beta(T^f)_{kl} u^l(z_2) A^f_\beta(z_2)\Big]\Big|0\Big>.
\end{align}
%Applying Wick's theorem, the quark fields are contracted to form light-quark propagators, while the gluon fields associated with the interaction vertices remain to be contracted with the gluonic EMT insertion. After performing the quark contractions, the interaction-induced correction to the up-quark propagator can be rewritten as
Using Wick's theorem, the quark fields are contracted into light-quark propagators,
\newpage
\begin{align}
&\Delta S_{u}^{ca',\ g}(y-x)=-\frac{g_s^2}{2}\int d^4z_1 d^4z_2
\Big<0\Big|	\mathcal{T}\Big[A^e_\alpha(z_1)A^f_\beta(z_2)T_{\mu\nu}^{g}(0)\Big]\Big|0\Big>
\nonumber\\[5pt]
&\qquad\quad\quad\times
\Bigg( S_u^{ci}(y-z_1) \gamma^\alpha (T^e)_{ij} S_u^{jk}(z_1-z_2)\gamma^\beta(T^f)_{kl} S_u^{la'}(z_2-x)
+ S_u^{ck}(y-z_2) \gamma^\beta (T^f)_{kl} S_u^{li}(z_2-z_1)\gamma^\alpha(T^e)_{ij} S_u^{ja'}(z_1-x)\Bigg).
\label{Su1}
\end{align}
To calculate the remaining gluon field contractions, the gluon fields in Eq.~\eqref{Su1} are written in terms of the gluon field strength tensor using the Fock-Schwinger gauge, $z^\mu A_\mu^a(z)=0$. In momentum space, the gluon field is given by \cite{Veliev:2010gb}
\allowdisplaybreaks
\begin{align}
A^e_\alpha(k)=-\frac{i}{2}(2\pi)^4G^e_{\alpha'\alpha}(0)\frac{\partial}{\partial k_{\alpha'}}\delta^{(4)}(k).
\label{eq:A}
\end{align}
Using the Fourier representation of the gluon field and substituting Eq.~\eqref{eq:A}, one obtains
\begin{align}
A^e_\alpha(z_1)=\frac{1}{2}z_1^{\alpha'}G^e_{\alpha'\alpha}(0).
\label{eq:A1}
\end{align}
The similar equation is used for the second gluon field $A_\beta^f(z_2)$ in Eq.~\eqref{Su1}.
By inserting Eq.~\eqref{eq:A1} into Eq.~\eqref{Su1} and using Eq.~\eqref{condensation gluon} for the vacuum expectation value of the gluon field strength, we obtain
\allowdisplaybreaks
\begin{align}
	\Delta S_{u}^{ca',\ g}(y-x)= \frac{g_s^2\ \big< G^2\big>}{96^2} \int d^4z_1 d^4z_2&\
	\Bigg[
	\frac{1}{8}\ S_u^{ci}(y-z_1) \big(z_{1\mu}\gamma_{\rho}-z_{1\rho}\gamma_{\mu}\big) (T^n)_{ij} S_u^{jk}(z_1-z_2)\big(z_2^\rho \gamma_{\nu}-z_{2 \nu}\gamma^\rho\big)(T^n)_{kl} S_u^{la'}(z_2-x)
	\nonumber\\[5pt]
	&+\frac{1}{8}\  S_u^{ci}(y-z_1) 	\big( z_1^\rho \gamma_{\nu}-z_{1 \nu}\gamma^\rho \big) (T^n)_{ij} S_u^{jk}(z_1-z_2)\big( z_{2 \mu}\gamma_{\rho}-z_{2 \rho} \gamma_{\mu}\big)(T^n)_{kl} S_u^{la'}(z_2-x)
	\nonumber\\[5pt]
	&+\frac{1}{8}\ S_u^{ck}(y-z_2) \big(z_2^\rho \gamma_{\nu}-z_{2 \nu}\gamma^\rho\big) (T^n)_{kl} S_u^{li}(z_2-z_1)\big(z_{1\mu}\gamma_{\rho}-z_{1\rho}\gamma_{\mu}\big)(T^n)_{ij} S_u^{ja'}(z_1-x)
	\nonumber\\[5pt]
	&+ \frac{1}{8}S_u^{ck}(y-z_2) \big( z_{2 \mu}\gamma_{\rho}-z_{2 \rho} \gamma_{\mu}\big) (T^n)_{kl} S_u^{li}(z_2-z_1)\big( z_1^\rho \gamma_{\nu}-z_{1 \nu}\gamma^\rho \big)(T^n)_{ij} S_u^{ja'}(z_1-x)
	\nonumber\\[5pt]
	&-\frac{ g_{\mu\nu}}{32}\ S_u^{ci}(y-z_1)\big(z_{1 \rho}\gamma_{\delta}-z_{1 \delta}\gamma_{\rho}\big) (T^n)_{ij} S_u^{jk}(z_1-z_2) \big(z_2^{ \delta}\gamma^{\rho}-z_{2}^\rho \gamma^\delta\big)(T^n)_{kl} S_u^{la'}(z_2-x)
	\nonumber\\[5pt]
	&-\frac{ g_{\mu\nu}}{32}\ S_u^{ci}(y-z_1)\big(z_1^{ \delta}\gamma^{\rho}-z_1^\rho \gamma^\delta\big) (T^n)_{ij} S_u^{jk}(z_1-z_2) \big(z_{2 \rho}\gamma_{\delta}-z_{2 \delta}\gamma_{\rho}\big)(T^n)_{kl} S_u^{la'}(z_2-x)
	\nonumber\\[5pt]
	&-\frac{ g_{\mu\nu}}{32}\ S_u^{ck}(y-z_2)\big(z_2^{ \delta}\gamma^{\rho}-z_{2}^\rho \gamma^\delta\big) (T^n)_{kl} S_u^{li}(z_2-z_1)\big(z_{1 \rho}\gamma_{\delta}-z_{1 \delta}\gamma_{\rho}\big)(T^n)_{ij} S_u^{ja'}(z_1-x)
	\nonumber\\[5pt]
	&-\frac{ g_{\mu\nu}}{32}\ S_u^{ck}(y-z_2)\big(z_{2 \rho}\gamma_{\delta}-z_{2 \delta}\gamma_{\rho}\big) (T^n)_{kl} S_u^{li}(z_2-z_1)\big(z_1^{ \delta}\gamma^{\rho}-z_1^\rho \gamma^\delta\big)(T^n)_{ij} S_u^{ja'}(z_1-x)
	\Bigg],
	\label{Delta S}
\end{align}
where $n$ denotes the color index of the gluon field.
The same procedure is followed for the other light-quark propagators in the correlation function to obtain the full contribution of the second class. The intermediate calculations are not shown here due to their length, and the final result is presented as
\allowdisplaybreaks
\begin{align}
\Pi_{\mu\nu}^{\text{QCD}, \ g}(p,p')&=i^2  \varepsilon_{abc}\varepsilon_{a'b'c'}\ \int d^4x\ e^{-i p\cdot x} \int d^4y \ e^{i p'\cdot y}\ \Pi_{\mu\nu}^{g}(x,y).
\label{Pi-QCD-second QCD}
\end{align}	
The explicit form of $\Pi_{\mu\nu}^{g}(x,y)$ includes all corrections to the light-quark propagators and is given in Eq.~\eqref{Pi2}.
Since the first class has no contribution to the correlation function, the total correlation function receives contributions only from the second class.

Using the light-quark propagator in Eq.~\eqref{propagator} and its correction in Eq.~\eqref{Delta S}, the correlation function on the QCD side is calculated from Eq.~\eqref{Pi2}, and the result is
\allowdisplaybreaks	
\begin{align}
\Pi_{\mu\nu}^{\text{QCD},\ g}(p,p')=\int d^4x\ e^{-i p\cdot x} \int d^4y \ e^{i p'\cdot y}\ &\Big[\Pi^{(\text{pert}),\ g}_{\mu\nu}(x,y)+\Pi^{(3D),\ g}_{\mu\nu}(x,y)+\Pi^{(4D),\ g}_{\mu\nu}(x,y)\nonumber\\[7pt]&
+\Pi^{(5D),\ g}_{\mu\nu}(x,y)+\Pi^{(6D),\ g}_{\mu\nu}(x,y)+\Pi^{(7D),\ g}_{\mu\nu}(x,y)+\cdots\Big],
\label{corrf2}
\end{align}	
here, $D$ represents the space-time dimension.
The operator product expansion of the correlation function on the QCD side contains a perturbative term, $\Pi_{\mu\nu}^{(\mathrm{pert}),\ g}$, and nonperturbative contributions from condensates with mass dimensions three to seven, denoted by $\Pi_{\mu\nu}^{(3D),\ g}$, $\Pi_{\mu\nu}^{(4D),\ g}$, $\Pi_{\mu\nu}^{(5D),\ g}$, $\Pi_{\mu\nu}^{(6D),\ g}$, and $\Pi_{\mu\nu}^{(7D),\ g}$, respectively.

The terms in Eq.~\eqref{corrf2} are first calculated in coordinate space and then transformed into momentum space using the Fourier transform \cite{Azizi:2017ubq},
\allowdisplaybreaks
\begin{align}
\dfrac{1}{(L^2)^{m_j}} =\int \frac{d^D p_j}{(2\pi)^D}e^{-ip_j\cdot L}\ i(-1)^{m_j+1}2^{D-2m_j}\pi^{D/2}
\frac{\Gamma[D/2-m_j]}{\Gamma[m_j]}\left(-\frac{1}{p_j^2}\right)^{D/2-m_j},
\label{fourie}
\end{align}
where $L$ denotes the separation between the points in the propagators.
The coordinate dependence in the correlation function is expressed in terms of derivatives with respect to the corresponding momenta. Specifically,
$y_\mu=-i\frac{\partial}{\partial p'_\mu}$, $x_\mu=i\frac{\partial}{\partial p_\mu}$, $z_{1\mu}=i\frac{\partial}{\partial q'_\mu}$, $z_{2\mu}=i\frac{\partial}{\partial q_\mu}
$.
Integrating over the coordinate variables $x$, $y$, $z_1$, and $z_2$ in $D$ dimensions produces Dirac delta functions in momentum space.
These delta functions are used to carry out the momentum integrations and reduce the number of independent momenta.
The remaining momentum integrals are evaluated using Feynman parametrization and a standard integral formula \cite{Azizi:2017ubq},
\allowdisplaybreaks
\begin{equation}
\int d^D\ell \frac{1}{(\ell^2 + \Delta)^n} = \dfrac{i \pi^{D/2} (-1)^{n}\Gamma[n-D/2]}{\Gamma[n] (-\Delta)^{n-D/2}}.
\label{integral}
\end{equation}
Using the dispersion relation, the invariant amplitudes on the QCD side can be written as double dispersion integrals,
\allowdisplaybreaks
\begin{align}
\Pi_{i}^{\mathrm{QCD}, \ g}(s_0,Q^2)=\int_{(2m_u+m_d)^2}^{s_0} ds\ \int_{(2m_u+m_d)^2}^{s_0} ds'\
\frac{\rho^g_i(s,s',Q^2)}{(s-p^2)(s'-p'^2)},
\label{spectral density}
\end{align}
where $\rho^g_i(s,s',Q^2)$ is the spectral density obtained from the corresponding invariant amplitude by separating the contributions of different operator dimensions in the OPE.
The continuum contribution is removed by restricting the integration regions in $s$ and $s'$ to the continuum threshold parameter $s_0$.
Since the initial and final states are both protons, the two dispersion integrals have the same integration limits.
The spectral densities are obtained from the imaginary parts of the corresponding invariant amplitudes on the QCD side,
\allowdisplaybreaks
\begin{align}
\rho^g_i(s,s',Q^2)=\frac{1}{\pi}\ \mathrm{Im}\left[\Pi_i^{\mathrm{QCD}, \ g}(s,s',Q^2)\right].
\end{align}
The imaginary parts are calculated using the following relation \cite{Azizi:2017ubq},
\allowdisplaybreaks
\begin{align}
\Gamma\left[\frac{D}{2}-n\right]\left(\frac{-1}{\Delta}\right)^{D/2-n}=\frac{(-1)^{n-1}}{(n-2)!}
(-\Delta)^{n-2}\ln[-\Delta],
\label{imarinarypart}
\end{align}
This relation identifies the logarithmic terms that contribute to the imaginary parts of the invariant amplitudes and to the spectral densities.
The calculations are carried out in $D=4$ space-time dimensions. Next, a double Borel transformation with respect to $p^2$ and $p'^2$ is applied to the invariant amplitudes on the QCD side, yielding the following expression for the Borel-transformed correlation function,
\allowdisplaybreaks
\begin{align}
\Pi_{\mu\nu}^{\mathrm{QCD},\ g}(s_0, M^2, Q^2)&=
\Pi_1^{\mathrm{QCD}, \ g}(Q^2)\ p_\mu p_\nu \mathbb{1}
+\Pi_2^{\mathrm{QCD}, \ g}(Q^2)\ p_\mu p'_\nu \mathbb{1}
+\Pi_3^{\mathrm{QCD}, \ g}(Q^2)\ p_\mu p_\nu \slashed{p}'
+\Pi_4^{\mathrm{QCD}, \ g}(Q^2)\ p'_\mu p'_\nu \slashed{p}
\nonumber\\&
+\Pi_{5}^{\mathrm{QCD}, \ g}(Q^2)\ g_{\mu\nu}\slashed{p}
+\Pi_{6}^{\mathrm{QCD}, \ g}(Q^2)\ p_\mu p_\nu \slashed{p}\slashed{p}'
+\Pi_{7}^{\mathrm{QCD}, \ g}(Q^2)\ p_\nu p'_\mu \slashed{p}\slashed{p}'
+\Pi_{8}^{\mathrm{QCD}, \ g}(Q^2)\ p_\mu p'_\nu \slashed{p}\slashed{p}'
\nonumber\\&
+\Pi_{9}^{\mathrm{QCD}, \ g}(Q^2)\ p'_\mu p'_\nu \slashed{p}\slashed{p}'
+\Pi_{10}^{\mathrm{QCD}, \ g}(Q^2)\ g_{\mu\nu}\slashed{p}\slashed{p}'
+\dots\ .
\label{eq:correlation ًQCD}
\end{align}
The invariant amplitudes $\Pi_i^{\mathrm{QCD},\ g}(Q^2)$ receive perturbative and nonperturbative contributions from the OPE. Their expressions are lengthy and are not presented here.
%The GFFs are determined by equating the coefficients of the independent Lorentz structures in the QCD side given in Eq.~\eqref{eq:correlation ًQCD} and the physical representation given in Eq.~\eqref{equation correlation physical}.
\subsubsection{QCD representation of the photon contribution}\label{QCD-photon}
In the photon sector, the first class of contractions involving the direct vacuum contraction of the EMT fields does not contribute because the vacuum expectation value of the electromagnetic EMT vanishes. Therefore, the photon contribution arises only from the interaction of the photon fields in the EMT operator with the light-quark fields through the quark-photon vertex.

Therefore, the light-quark propagator in the correlation function contains the correction due to the photon,
\allowdisplaybreaks
\begin{align}
S^{ab}_\psi(y-x)\;\Longrightarrow\;\Delta S^{ab,\ \gamma}_\psi(y-x).
\end{align}
The photon-induced correction to the light-quark propagator is obtained using the same Green's function formalism described in Eq.~\eqref{Green function}, with the gluonic EMT insertion $T_{\mu\nu}^{g}(0)$ replaced by the electromagnetic EMT insertion $T_{\mu\nu}^{\gamma}(0)$, and the quark--gluon interaction Lagrangian replaced by the corresponding quark--photon interaction term,
\allowdisplaybreaks
\begin{align}
\mathcal{L}^{q\gamma}_{\mathrm{int}}(z)=-e_q\ \bar q^{\ a}(z)\gamma^\beta q^{\ a}(z)A_\beta(z),
\label{lagrangian interaction photon}
\end{align}
where $e_q=Q_q e$ is the quark electric charge, with $Q_u=2/3$ and $Q_d=-1/3$.
Following the procedure in Subsection~\ref{QCD-gluon} and using the quark--photon interaction Lagrangian in Eq.~\eqref{lagrangian interaction photon}, the correction to the light-quark propagator due to the photon is obtained.
As an example, the correction to the up-quark propagator $S_u^{ca'}(y-x)$ is
\allowdisplaybreaks
\begin{align}
\Delta S_{u}^{ca',\ \gamma}(y-x)=-\frac{e_u^2}{2}\int d^4z_1 d^4z_2\ \Big\langle 0 \Big|\mathcal{T}
\Big[u^c(y)T_{\mu\nu}^{\gamma}(0)\bar{u}^{a'}(x)\bar{u}^{i}(z_1)\gamma^\alpha u^{i}(z_1) A_\alpha(z_1)
\bar{u}^{j}(z_2)\gamma^\beta u^{j}(z_2) A_\beta(z_2)\Big]\Big|0 \Big\rangle.
\end{align}
Applying Wick's theorem, the quark fields are contracted to form light-quark propagators,
\allowdisplaybreaks
\begin{align}
\Delta S_{u}^{ca',\ \gamma}(y-x)&=-\frac{e_u^2}{2}\int d^4z_1 d^4z_2\Big\langle 0\Big|\mathcal{T}
\Big[A_\alpha(z_1)A_\beta(z_2)T_{\mu\nu}^{\gamma}(0)\Big]\Big|0\Big\rangle
\nonumber\\[5pt]&~~~~~~~~~~~\times
\Bigg(S_u^{ci}(y-z_1)\gamma^\alpha S_u^{ij}(z_1-z_2)\gamma^\beta S_u^{ja'}(z_2-x)
+S_u^{ci}(y-z_2)\gamma^\beta S_u^{ij}(z_2-z_1)\gamma^\alpha S_u^{ja'}(z_1-x)\Bigg).
\label{Su1,photon}
\end{align}
The vacuum expectation value in Eq.~\eqref{Su1,photon} is evaluated using the electromagnetic EMT in Eq.~\eqref{field strength tensor-photon} and Wick's theorem. We introduce the following shorthand notation:
\allowdisplaybreaks
\begin{align}
G_{\alpha;\ \mu\rho}(z_1)=\left\langle 0\left|\mathcal{T}\left[A_\alpha(z_1)F_{\mu\rho}(0)\right]
\right|0\right\rangle=-\Big[\partial_\mu D^{(0)}_{\alpha\rho}(z_1)-\partial_\rho D^{(0)}_{\alpha\mu}(z_1)\Big],
\label{pro-photon}
\end{align}
where $D_{\alpha\beta}^{(0)}(z)$ denotes the free photon propagator.The derivatives in the above relation are initially taken with respect to the coordinate of the field-strength tensor at the EMT insertion point. By translational invariance of the photon propagator, they can be written as derivatives with respect to $z_1$. Accordingly, the photon contractions are evaluated using the free photon propagator. Substituting Eq.~\eqref{pro-photon} into Eq.~\eqref{Su1,photon}, the correction to the light-quark propagator due to the photon is
\allowdisplaybreaks
\begin{align}
\Delta S_{u}^{ca',\ \gamma}(y-x)&=-\frac{e_u^2}{2}\int d^4z_1 d^4z_2
\Bigg(S_u^{ci}(y-z_1)\gamma^\alpha S_u^{ij}(z_1-z_2)\gamma^\beta S_u^{ja'}(z_2-x)
+S_u^{ci}(y-z_2)\gamma^\beta S_u^{ij}(z_2-z_1)\gamma^\alpha S_u^{ja'}(z_1-x)\Bigg)
\nonumber\\[5pt]&~~~~~\times
\Bigg(-\Big[G_{\alpha;\ \mu\rho}(z_1)G_{\beta;\ \nu}^{\ \ \rho}(z_2)
+G_{\alpha;\ \nu}^{\ \ \rho}(z_1)G_{\beta;\ \mu\rho}(z_2)\Big]+\frac{1}{4}g_{\mu\nu}
\Big[G_{\alpha;\ \rho\sigma}(z_1)G_{\beta}^{\ \rho\sigma}(z_2)+
G_{\alpha;\ \rho\sigma}(z_2)G_{\beta}^{\ \rho\sigma}(z_1)\Big]\Bigg).
\label{Su1,photon1}
\end{align}
Using the Fourier representation of the free photon propagator and the Fourier transforms of the coordinate-space light-quark propagators, the correction is written in momentum space. 
The result is given in Eq.~\eqref{Su1-photon-momentum}. The same procedure is applied to the remaining light-quark propagators to obtain the complete photon contribution to the correlation function. The final result on the QCD side is
\allowdisplaybreaks
\begin{align}
\Pi_{\mu\nu}^{\text{QCD},\ \gamma}(p,p')&=i^2  \varepsilon_{abc}\varepsilon_{a'b'c'}\ \int d^4x\ e^{-i p\cdot x} \int d^4y \ e^{i p'\cdot y}\ \Pi_{\mu\nu}^\gamma,
\label{Pi-QCD side-photon}
\end{align}	
The result is given in Eq.~\eqref{Pigamma}. Using the light-quark propagators, the correlation function is separated into the perturbative contribution and the nonperturbative contributions from operators of different dimensions, up to dimension seven.

To evaluate the correlation function, the integrations over the coordinate variables $x$, $y$, $z_1$, and $z_2$ are performed. These integrations produce momentum-conserving delta functions that simplify the momentum integrals. The remaining momentum integrals are evaluated using the Feynman parametrization in Eq.~\eqref{integral}.
Following the procedure used for the gluon contribution, the invariant amplitudes for the photon contribution are expressed in terms of the corresponding spectral densities. These amplitudes are written as double dispersion integrals,
\allowdisplaybreaks
\begin{align}
	\Pi_{i}^{\mathrm{QCD},\ \gamma}(s_0,Q^2)
	=
	\int_{(2m_u+m_d)^2}^{s_0} ds
	\int_{(2m_u+m_d)^2}^{s_0} ds'\
	\frac{\rho_i^\gamma(s,s',Q^2)}
	{(s-p^2)(s'-p'^2)}.
\end{align}
where $\rho_i^\gamma(s,s',Q^2)$ denotes the photon spectral density obtained from the corresponding invariant amplitude after separating the different operator dimension contributions in the OPE. The spectral densities are obtained from the imaginary parts of these amplitudes,
$\rho_i^\gamma(s,s',Q^2)=\mathrm{Im}\big[\Pi_i^{\mathrm{QCD},\ \gamma}(s,s',Q^2)\big]/{\pi}$.
Using Eq.~\eqref{imarinarypart}, the imaginary parts of the invariant amplitudes for the photon contribution are obtained.
The calculations are performed in four dimensional space-time, and the photon contribution to the correlation function in the Borel scheme is obtained as
\allowdisplaybreaks
\begin{align}
	\Pi_{\mu\nu}^{\mathrm{QCD},\ \gamma}(s_0, M^2, Q^2)&=
	\Pi_1^{\mathrm{QCD}, \ \gamma}(Q^2)\ p_\mu p_\nu \mathbb{1}
	+\Pi_2^{\mathrm{QCD}, \ \gamma}(Q^2)\ p_\mu p'_\nu \mathbb{1}
	+\Pi_3^{\mathrm{QCD}, \ \gamma}(Q^2)\ p_\mu p_\nu \slashed{p}'
	+\Pi_4^{\mathrm{QCD}, \ \gamma}(Q^2)\ p'_\mu p'_\nu \slashed{p}
	\nonumber\\&
	+\Pi_{5}^{\mathrm{QCD}, \ \gamma}(Q^2)\ g_{\mu\nu}\slashed{p}
	+\Pi_{6}^{\mathrm{QCD}, \ \gamma}(Q^2)\ p_\mu p_\nu \slashed{p}\slashed{p}'
	+\Pi_{7}^{\mathrm{QCD}, \ \gamma}(Q^2)\ p_\nu p'_\mu \slashed{p}\slashed{p}'
	+\Pi_{8}^{\mathrm{QCD}, \ \gamma}(Q^2)\ p_\mu p'_\nu \slashed{p}\slashed{p}'
	\nonumber\\&
	+\Pi_{9}^{\mathrm{QCD}, \ \gamma}(Q^2)\ p'_\mu p'_\nu \slashed{p}\slashed{p}'
	+\Pi_{10}^{\mathrm{QCD}, \ \gamma}(Q^2)\ g_{\mu\nu}\slashed{p}\slashed{p}'
	+\dots\ .
	\label{eq:correlation ًQCD-photon}
\end{align}
The invariant amplitudes $\Pi_i^{\mathrm{QCD},\ \gamma}(Q^2)$ have lengthy expressions, and their explicit forms are not shown here.

The GFFs associated with the gluon and photon sectors are obtained separately by matching the coefficients of the independent Lorentz structures in the corresponding QCD side correlation functions in Eqs.~\eqref{eq:correlation ًQCD} for the gluon and \eqref{eq:correlation ًQCD-photon} for the photon, with those in the physical representation of the correlation function in Eq.~\eqref{equation correlation physical}.

\section{NUMERICAL ANALYSES}\label{sec:numerical} 
Using the QCDSR framework, we perform a numerical analysis of the gluon and photon GFFs of the proton. The input parameter values used in the calculations are listed in Table~\ref{input}.
\begin{table}[!htb]
	\centering
	\begin{minipage}{0.96\textwidth} % slightly wider
		\centering
		\renewcommand{\arraystretch}{2} % increases row height
		\setlength{\tabcolsep}{12pt}      % increases column padding
		\begin{tabular}{|c|c|c|c|c|c|}
			\hline
			\text{Input parameters} & \text{Values} & \text{Input parameters} & \text{Values} \\
			\hline\hline
			$m_u$ & $0.00216\pm0.00007\ \text{GeV}$ \cite{ParticleDataGroup:2024cfk}  & $\alpha_s$ &$(0.118 \pm 0.005)$ \cite{DELPHI:1993ukk} \\
			\hline
			$m_d$ & $0.00470\pm0.00004\ \text{GeV}$ \cite{ParticleDataGroup:2024cfk} & $g_s$& $\sqrt{4\pi \alpha_s}$ \\
			\hline
			$m_{N}$ & $ 0.93827208943\pm 0.00000000029 \ \text{GeV}$  \cite{ParticleDataGroup:2024cfk} & $\langle \frac{\alpha_s}{\pi} G^2 \rangle $ & $(0.012\pm0.004)$ $~\mathrm{GeV}^4 $ \cite{Belyaev:1982cd}  \\
			\hline
			$\left\langle \bar{q}q \right\rangle,\ $\text{with}\ ($q=u,d$) & $(-0.24\pm 0.01)^3$ $\mathrm{GeV}^3$ \cite{Belyaev:1982sa} & $\left\langle \bar{q} g_s \sigma G q\right\rangle$ & $(0.8 \pm 0.1)\left\langle \bar{q}q \right\rangle \ \mathrm{GeV}^5$ \cite{Belyaev:1982sa}\\
			\hline
			$e_u$ & $0.2019$ & $e_d$ & $-0.1009$
			\\
			\hline
		\end{tabular}
	\end{minipage}
	\caption{Input parameters used in the numerical analysis and their corresponding values.}
	\label{input}
\end{table}

The numerical analysis requires suitable working intervals for the auxiliary parameters $M^2$, $s_0$, and $\beta$ in the QCDSR formulation. The Borel parameter $M^2$ is chosen to ensure good convergence of the OPE series and a sufficiently large contribution from the ground state pole. The continuum threshold parameter $s_0$ is set to separate the proton ground-state contribution from higher excited states and the continuum. The following criteria are used to ensure reliable sum rules \cite{Aliev:2016jnp},
\allowdisplaybreaks
\begin{align}
	\mathrm{PC}(Q^2)
	&=
	\frac{\Pi_{i}^{\mathrm{QCD}}(s_0,M^2,Q^2)}
	{\Pi_{i}^{\mathrm{QCD}}(\infty,M^2,Q^2)}
	\geq 0.5,
	\qquad
	\mathrm{R}(M^2,Q^2)
	=
	\frac{\Pi_{i}^{\mathrm{QCD},\ \mathrm{Dim}\ D_{\mathrm{max}}}
		(s_0,M^2,Q^2)}
	{\Pi_{i}^{\mathrm{QCD}}(\infty,M^2,Q^2)}
	\leq 0.08,
\end{align}
here, $i$ denotes the selected Lorentz structures, and $D_{\mathrm{max}}$ corresponds to the highest operator dimension included in the OPE calculation. The upper and lower bounds of the Borel parameter are determined by the pole contribution and OPE convergence criteria, respectively. Based on these criteria and previous QCDSR analyses of the proton, the following working intervals are used \cite{Dehghan:2025ncw, Aliev:2019tmk, Azizi:2020jog, Olamaei:2023bpi},
\allowdisplaybreaks
\begin{align}
	1.2~\mathrm{GeV}^2 \leq M^2 \leq 2.2~\mathrm{GeV}^2,
	\qquad
	2.25~\mathrm{GeV}^2 \leq s_0 \leq 2.35~\mathrm{GeV}^2 .
\end{align}
Moreover, the working range of the mixing parameter $\beta$ is determined from the stability of the extracted GFFs and is found to be $\beta=-2\pm0.4$ \cite{Dehghan:2025ncw}.
The dependence of the gluon and photon GFFs on the Borel parameter $M^2$ has been investigated at $Q^2=1~\mathrm{GeV}^2$ for three representative values of the continuum threshold, $s_0=2.25$, $2.30$, and $2.35~\mathrm{GeV}^2$. We observe that the extracted GFFs remain stable over the selected Borel parameter range.

\begin{figure}[!htb]
	\centering
	\includegraphics[width=0.42\textwidth]{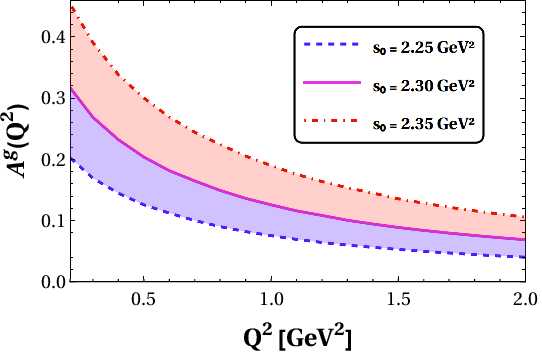}~~~~~~~~
	\includegraphics[width=0.42\textwidth]{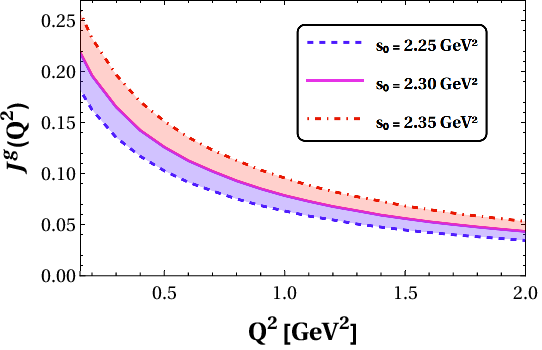}
	
	\vspace{0.1cm}
	\includegraphics[width=0.42\textwidth]{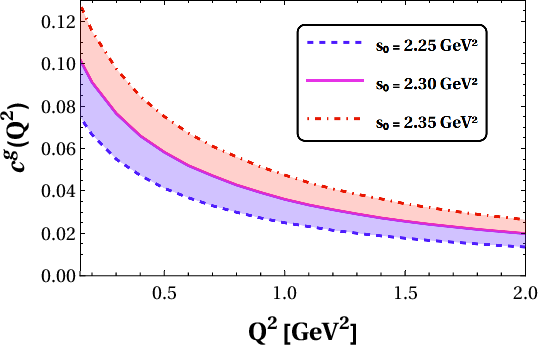}~~~~~~~~
	\includegraphics[width=0.42\textwidth]{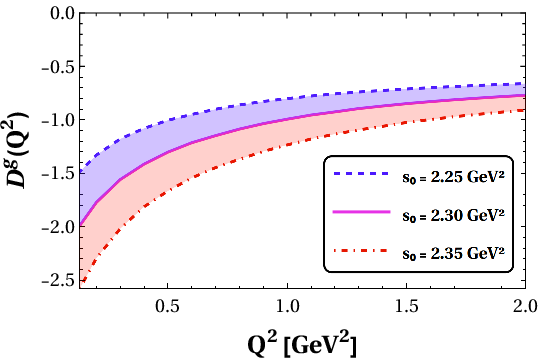}
	
	\caption{The $Q^2$ dependence of the gluon GFFs of the proton at $M^2 = 1.7~\text{GeV}^2$ for three values of the continuum threshold $s_0$.}
	\label{Qgluon}
\end{figure}

\begin{figure}[!htb]
	\centering
	\includegraphics[width=0.42\textwidth]{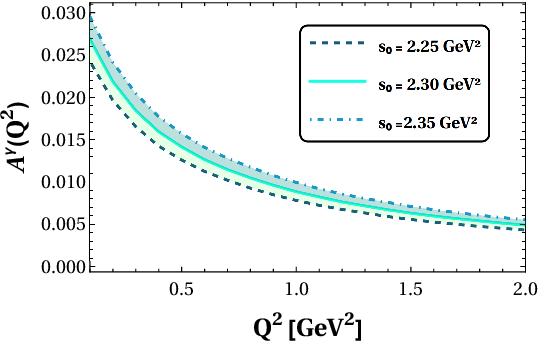}~~~~~~~~
	\includegraphics[width=0.42\textwidth]{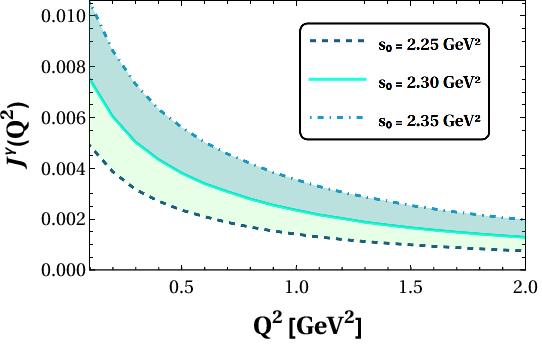}
	
	\vspace{0.1cm}
	\includegraphics[width=0.42\textwidth]{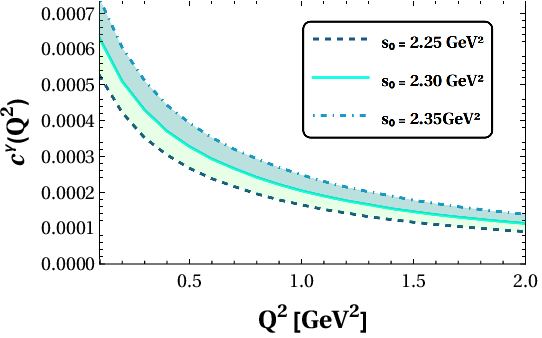}~~~~~~~~
	\includegraphics[width=0.42\textwidth]{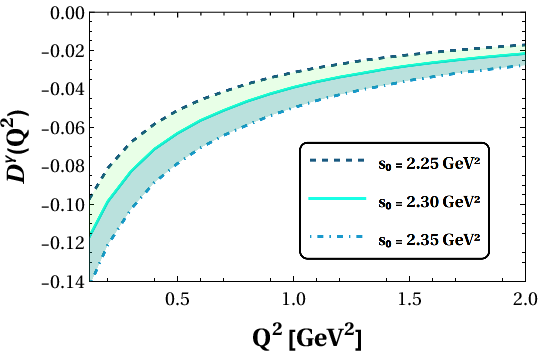}
	\caption{The $Q^2$ dependence of the photon GFFs of the proton at $M^2 = 1.7~\text{GeV}^2$ for three values of the continuum threshold $s_0$.}
	\label{Qphoton}
\end{figure}

The \(Q^2\) dependence of the gluon and photon GFFs is shown in Figs.~\ref{Qgluon} and \ref{Qphoton}. The results are obtained for three values of the continuum threshold, \(s_0=2.25\), \(2.30\), and \(2.35~\mathrm{GeV}^2\), with the Borel parameter fixed at \(M^2=1.7~\mathrm{GeV}^2\). All GFFs decrease smoothly as the momentum transfer increases.
To describe the $Q^2$ dependence of the gluon and photon GFFs, we employ the $\mathbf{p}$-pole form \cite{Mamo:2022eui, Pefkou:2021fni}, given by
\allowdisplaybreaks
\begin{align}
{\cal F}(Q^2)=\frac{{\cal F}(0)}{\Bigg(1+\dfrac{Q^2}{m_\mathbf{p}^2} \Bigg)^\mathbf{p} },
	\label{eq:FitFun}
\end{align}
where $\mathcal{F}(0)$ represents the value of the form factor at zero momentum transfer, $m_\mathbf{p}$ denotes an effective mass parameter with units of GeV, and $\mathbf{p}$ is a dimensionless parameter that determines the power behavior of the $Q^2$ dependence.
The fitted parameters obtained from this analysis are listed in Table~\ref{table:fitparameters}.
As shown in Figs.~\ref{Qgluon} and \ref{Qphoton}, the \(\mathbf{p}\)-pole parametrization describes the QCDSR results well over the considered momentum transfer region.
\begin{table}[t]
	\centering
	\begin{minipage}[t]{0.492\textwidth}
		\centering
		\renewcommand{\arraystretch}{1.5}
		\setlength{\tabcolsep}{10pt}
		
		\begin{tabular}{|c|c|c|c|}
			\hline
			gluon GFF & ${\cal F}(0)$ & $m_\mathbf{p}$ & $\mathbf{p}$ \\
			\hline\hline
			$A^{g}(Q^2)$ & $0.46\pm0.06$ & $0.52\pm0.09$ & $0.69\pm0.06$\\
			\hline
			$J^{g}(Q^2)$ & $0.26\pm0.08$ & $1.16\pm0.05$ & $1.75\pm0.03$\\
			\hline
			$\bar{c}^{g}(Q^2)$ & $0.13\pm0.03$ & $0.64\pm0.10$ & $0.91\pm0.03$\\
			\hline
			$D^{g}(Q^2)$ & $-2.54\pm0.12$ & $1.69\pm0.08$ & $1.93\pm0.06$\\
			\hline
		\end{tabular}
	\end{minipage}
	\hfill
	\begin{minipage}[t]{0.492\textwidth}
		\centering
		\renewcommand{\arraystretch}{1.5}
		\setlength{\tabcolsep}{10pt}
		
		\begin{tabular}{|c|c|c|c|}
			\hline
			photon GFF & ${\cal F}(0)$ & $m_\mathbf{p}$ & $\mathbf{p}$ \\
			\hline\hline
			$A^{\gamma}(Q^2)$ & $0.03\pm0.01$ & $0.39\pm0.02$ & $0.63\pm0.03$\\
			\hline
			$J^{\gamma}(Q^2)$ & $0.01\pm0.01$ & $1.15\pm0.05$ & $1.75\pm0.02$\\
			\hline
			$\bar{c}^{\gamma}(Q^2)$ & $0.00\pm0.00$ & $0.24\pm0.04$ & $0.45\pm0.04$\\
			\hline
			$D^{\gamma}(Q^2)$ & $-0.14\pm0.02$ & $1.26\pm0.06$ & $1.90\pm0.04$\\
			\hline
		\end{tabular}
	\end{minipage}
	
	\caption{Fit parameters obtained from the $\mathbf{p}$-pole parametrization of the gluon and photon gravitational form factors of the proton.}
		\label{table:fitparameters}
\end{table}
The results presented in Table~\ref{table:fitparameters} show a clear distinction between the gluon and photon contributions to the proton GFFs. 
In general, the gluon GFFs have larger magnitudes than the corresponding photon GFFs, reflecting the different contributions of these two sectors to the proton EMT. Furthermore, the gluon and photon GFFs show different $Q^2$ dependences.

We compare our results with previous findings in the literature. 
For the photon sector, direct comparison with previous calculations of photon GFFs is not possible because only a few studies are available.
Nevertheless, the obtained value of $A^\gamma(0)$ has the same order of magnitude as the photon momentum fraction of the proton reported in recent photon PDF analyses, consistent with the small electromagnetic contribution to the proton EMT \cite{Manohar:2017eqh, NNPDF:2024djq}. 
For the gluon sector, a comparison with previous theoretical studies is possible. The corresponding values of the gluon GFFs obtained using different theoretical approaches, including lattice QCD, holographic QCD, phenomenological analyses, and light-front approaches, are summarized in Table~\ref{tab:comparison}.
\begin{table}[!htb]
	\centering
	\renewcommand{\arraystretch}{1.3}
	\setlength{\tabcolsep}{8pt}
	\begin{tabular}{|c|c|c|c|c|}
		\hline
		Approach & $A^g(0)$ & $J^g(0)$ & $\bar{c}^g(0)$ & $D^g(0)$\\
		\hline\hline
		This work (QCDSR) 
		& $0.46\pm0.16$ 
		& $0.26\pm0.08$ 
		& $0.13\pm0.08$ 
		& $-2.54\pm0.82$\\
		\hline
		
		Lattice QCD (dipole fit) \cite{Hackett:2023rif}
		& $0.50\pm 0.27$ 
		& $0.25\pm 0.13$ 
		& -- 
		& $-2.57\pm 0.84$\\
		\hline
		
		Lattice QCD (dipole fit) \cite{Shanahan:2018pib}
		& $0.58$ 
		& -- 
		& -- 
		& $-10.00$\\
		\hline
		
		Lattice QCD (tripole fit) \cite{Pefkou:2021fni}
		& $0.43\pm 0.39$ 
		& $0.26\pm 0.26$ 
		& -- 
		& $-1.93\pm 0.53$\\
		\hline
		
		Lattice QCD ($z$-expansion fit) \cite{Hackett:2023rif}
		& $0.53\pm 0.31$ 
		& $0.23\pm 0.27$ 
		& -- 
		& $-2.15\pm 0.32$\\
		\hline
		
		Lattice QCD ($z$-expansion fit) \cite{Pefkou:2021fni}
		& $0.41\pm 0.41$ 
		& $0.21\pm 0.57$ 
		& -- 
		& $-0.40\pm 1.20$\\
		\hline
		
		Holographic QCD  \cite{Mamo:2022eui}
		& $0.43$ 
		& -- 
		& -- 
		& $-0.40$\\
		\hline

		Phenomenological analysis ($J/\psi$ production) \cite{Guo:2023pqw}
		& $0.41$ 
		& -- 
		& -- 
		& $-1.49\pm 0.27$\\
		\hline
		
		Light-front approach  \cite{Tandy:2025tea}
		& $0.25$ 
		& $0.21$
		& $0.08$ 
		& --\\
		\hline
		
	\end{tabular}
	\caption{Comparison of the gluon GFFs of the proton obtained in this work with previous results from different approaches.}
	\label{tab:comparison}
\end{table}
\newpage

\vspace{0.5cm}
\newpage
For the results reported in Refs.~\cite{Shanahan:2018pib, Mamo:2022eui, Guo:2023pqw, Tandy:2025tea}, only the central values are presented, since the corresponding uncertainties were not provided in the original works. The entries denoted by ``--'' indicate that the corresponding form factors were not reported in the original studies.

The comparison presented in Table~\ref{tab:comparison} shows that the results obtained in this work are in good agreement with previous results reported in the literature, despite the different approaches employed in their determination. In particular, the obtained value of $A^g(0)$ agrees well with previous results, indicating a consistent determination of the gluon contribution to the proton momentum. Similarly, the obtained value of $J^g(0)$ is consistent with previous results, indicating a compatible determination of the gluon contribution to the proton angular momentum. In contrast, the $D$-term shows larger variations among the available results, which may be related to its sensitivity to the details of the determination procedure, while remaining qualitatively consistent across different approaches.

Beyond the comparison of the GFFs, these form factors also provide information on the spatial properties of the gluon and photon contributions to the proton. The corresponding mass and scalar radii associated with the gluon and photon contributions are calculated from the obtained GFFs as follows:
\allowdisplaybreaks
\begin{align}
&\langle r_m^2\rangle_j=\frac{1}{A^j(0)+\bar c^j(0)}
\Bigg[6\frac{d A^j(t)}{dt}\Big|_{t=0}+6\frac{d \bar c^j(t)}{dt}\Big|_{t=0}
-\frac{3}{2m^2}\Big(A^j(0)-2J^j(0)+D^j(0)\Big)\Bigg],
\nonumber\\[5pt]
&
\langle r_s^2\rangle_j=\frac{1}{A_j(0)+4\bar{c}_j(0)}
\Bigg[6\left.\frac{dA_j(t)}{dt}\right|_{t=0}+24\left.\frac{d\bar{c}_j(t)}{dt}\right|_{t=0}
-\frac{3}{2m^2}\Big(A_j(0)-2J_j(0)+3D_j(0)\Big)\Bigg].
\label{radii-j}
\end{align}
A detailed derivation of these relations is given in Appendix~\ref{app:radii}.
Using Eq.~\eqref{radii-j} and the obtained GFFs, we find the following numerical values for the mass and scalar radii:
\allowdisplaybreaks
\begin{align}
& \langle r_m\rangle_g=0.934\pm 0.094\ \text{fm},\ \qquad \langle r_s\rangle_g=1.037\pm 0.075\ \text{fm}
\nonumber\\[7pt]
& \langle r_m\rangle_\gamma=1.120\pm 0.082\ \text{fm},\ \qquad \langle r_s\rangle_\gamma=1.369\pm 0.126\ \text{fm}.
\end{align}
The smaller gluon mass and scalar radii suggest that the gluon contributions are more spatially localized than the corresponding photon contributions. This difference may be related to the distinct nature of the strong and electromagnetic interactions.

The photon mass and scalar radii are obtained from the fitted photon GFFs in the range \(0<Q^2<2~\mathrm{GeV}^2\). However, the long range electromagnetic field must be considered when interpreting these radii. As discussed in Ref.~\cite{Freese:2022jlu} for the electron, this field can lead to infrared singularities in the mass radius. A similar issue may arise for the proton because of its long range Coulomb field. This contribution is not treated separately in the present calculation. Therefore, the finite photon mass radius obtained from the multipole fit should not be taken as the radius of the full electromagnetic energy distribution. The photon scalar radius should also be interpreted with care.

\begin{figure}[!htb]
	\centering
	
	\begin{minipage}{0.48\textwidth}
		\centering
		\includegraphics[width=\textwidth]{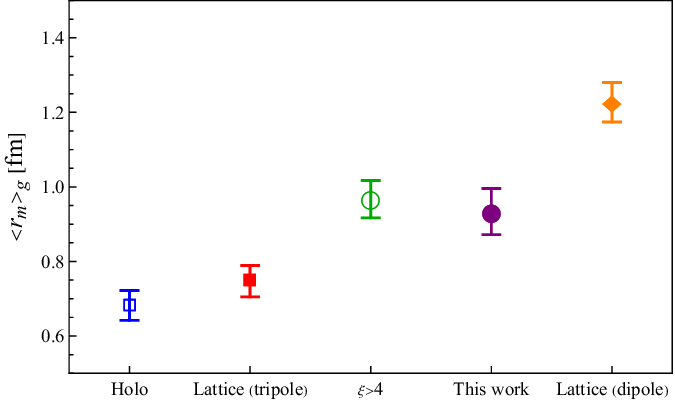}
	\end{minipage}
	\hfill
	\begin{minipage}{0.48\textwidth}
		\centering
		\includegraphics[width=\textwidth]{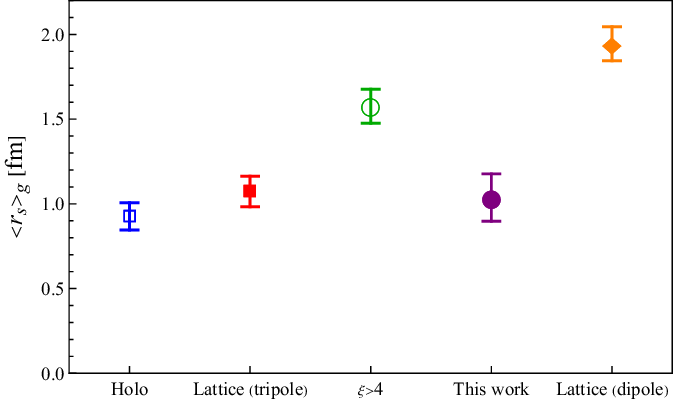}
	\end{minipage}
	
	\caption{Comparison of the gluon mass (left) and scalar (right) radii obtained in this work with previous results from different approaches, including lattice QCD using a dipole fit \cite{Shanahan:2018pib}, lattice QCD using a tripole fit \cite{Pefkou:2021fni}, holographic QCD \cite{Mamo:2022eui}, and the $\zeta>4$ approach \cite{Guo:2023pqw}.}
	\label{fig:radii}
\end{figure}

For the photon contribution, no direct comparison with previous results is provided due to the limited availability of corresponding determinations in the literature. For the gluon contribution, the obtained radii can be compared with the corresponding results reported in the literature. The comparison is presented in Fig.~\ref{fig:radii}.

The mass and scalar radii provide complementary information on the spatial structure of the gluonic contribution to the proton. The mass radius characterizes the spatial distribution of the gluon energy contribution, while the scalar radius is associated with the scalar part of the gluon EMT and provides information on the related mechanical structure. The difference between these two radii originates from the different combinations of gluon GFFs entering their definitions. 

Figure~\ref{fig:radii} compares the gluon mass and scalar radii obtained in this work with the available results from other approaches. The values obtained in this work fall within the range of the available predictions for both radii, although their relative positions within these ranges differ.

%These results indicate that the spatial distribution of the gluon energy inside the proton is relatively stable across different approaches, whereas the scalar structure shows a stronger dependence on the details of the employed determinations, consistent with its sensitivity to the $D$-term contribution.

\section{SUMMARY AND CONCLUSION}\label{sec:conclusion} 
The GFFs associated with the individual contributions to the proton EMT provide fundamental information about the internal structure of the proton, and their determination remains one of the central challenges in hadron physics. In this work, we have investigated the gluon and photon contributions to the gravitational structure of the proton within the QCDSR framework. This study provides an independent QCDSR investigation of these sectors and complements previous studies of the proton gravitational structure.

The correlation function was constructed from both the physical and QCD representations. On the physical side, the matrix element of the EMT was decomposed into independent Lorentz structures parameterized by the GFFs. On the QCD side, the gluon contribution was evaluated by taking into account both the vacuum expectation value of the gluon field-strength tensors and the quark--gluon interaction effects included through the corresponding correction to the quark propagator. The photon contribution was incorporated through the quark--photon interaction effects entering via the corresponding quark propagator correction.

Since the present study focuses only on the gluon and photon contributions to the proton EMT, without including the quark sector, these two sectors are described by four independent GFFs. This allows the corresponding non-conserved form factor contribution, represented by $\bar c^j(Q^2)$, to be consistently taken into account in the determination of the proton gravitational structure.

The obtained gluon GFFs provide insight into the energy and angular momentum distributions carried by the gluons inside the proton. The values of the gluon energy and angular momentum form factors at zero momentum transfer are found to be consistent with previous theoretical results, supporting the extraction of the gluon contribution to the proton momentum and spin structure. In contrast, the $D$-term shows a stronger sensitivity to different theoretical approaches, reflecting its dependence on the details of the nonperturbative dynamics and the extraction methodology.

Furthermore, the non-conserved form factor $\bar c^g(Q^2)$ has been explicitly retained in the present analysis due to the separate treatment of the gluon sector. This leads to a more complete description of the gluon contribution to the proton EMT without imposing the constraints that arise when the quark and gluon contributions are combined.
	
The photon contribution to the proton EMT is found to be much smaller than the gluon contribution, reflecting the subleading role of the electromagnetic sector in the proton gravitational structure. Nevertheless, its explicit determination provides complementary information on the electromagnetic contribution to the proton EMT.	

Beyond the $Q^2$-dependence of the GFFs, we further investigated the spatial structure of the gluon and photon contributions through the corresponding mass and scalar radii of the proton. The obtained gluon mass radius is found to be consistent with previous theoretical predictions, indicating a relatively stable spatial distribution of the gluon energy inside the proton. In contrast, the gluon scalar radius exhibits a larger variation among different approaches due to its stronger sensitivity to the $D$-term contribution. The calculated photon radii provide an estimate of the spatial extent of the electromagnetic contribution to the proton EMT.

The results obtained in this work show that the QCDSR framework provides a consistent approach for investigating the separate gluon and photon contributions to the gravitational structure of the proton. These results can serve as a basis for further investigations of the proton gravitational structure and more comprehensive studies of the QCD energy momentum tensor.
	
\section*{ACKNOWLEDGMENTS} 
K. Azizi thanks Iran national science foundation (INSF) for the partial financial support supplied under the elites Grant No. 40405095. 
	
	\section*{Declaration of AI-assisted language editing}

The authors have used AI tools, only for language editing and improving the readability of the manuscript. The authors have reviewed and edited the manuscript and take full responsibility for its scientific content.
	
\appendix
\renewcommand{\thesection}{\Alph{section}}
\renewcommand{\thesubsection}{\thesection. \arabic{subsection}}	
\section{The functions of physical side} \label{appA}	
In this appendix, we present the invariant amplitudes $\Pi_i^{\mathrm{phy},\j}(Q^2)$ obtained from the physical representation of the correlation function. Their dependence on the GFFs and the relevant physical parameters of the proton is explicitly shown as follows:	
\allowdisplaybreaks
\begin{align}
	\begin{array}{l@{\hspace{2cm}}l@{\hspace{2cm}}l} 
		\Pi_1^{\mathrm{phy}, \ j}(Q^2)=\dfrac{m}{4}\ \Big(A^j(Q^2)+D^j(Q^2)\Big), 
		&
		\Pi_{6}^{\mathrm{phy}, \ j}(Q^2)=\dfrac{1}{4m}\ \Big(A^j(Q^2)+D^j(Q^2)-J^j(Q^2)\Big),
		\\
		\\
		\Pi_2^{\mathrm{phy}, \ j}(Q^2)=\dfrac{m}{4}\ \Big(A^j(Q^2)-D^j(Q^2)\Big),
		&
		\Pi_{7}^{\mathrm{phy}, \ j}(Q^2)=\dfrac{1}{4m}\ \Big(A^j(Q^2)-D^j(Q^2)-J^j(Q^2)\Big),
		\\
		\\
		\Pi_3^{\mathrm{phy}, \ j}(Q^2)=\dfrac{1}{4}\ \Big(A^j(Q^2)+D^j(Q^2)\Big),
		&
		\Pi_8^{\mathrm{phy}, \ j}(Q^2)=\dfrac{1}{4m}\ \Big(A^j(Q^2)-D^j(Q^2)-2J^j(Q^2)\Big),
\\
\\
		\Pi_4^{\mathrm{phy}, \ j}(Q^2)=\dfrac{1}{4}\ \Big(A^j(Q^2)-D^j(Q^2)\Big),
		&
		\Pi_9^{\mathrm{phy}, \ j}(Q^2)=\dfrac{1}{4m}\ \Big(A^j(Q^2)+D^j(Q^2)-2J^j(Q^2)\Big),
		\\
		\\
		\Pi_5^{\mathrm{phy}, \ j}(Q^2)=m^2 \bar{c}^j(Q^2)-\dfrac{1}{4}D^j(Q^2),
		&
		\Pi_{10}^{\mathrm{phy}, \ j}(Q^2)=m\bar{c}^j(Q^2)-\dfrac{1}{4m}D^j(Q^2).
	\end{array}
	\label{function omega}
\end{align}

\section{QCD  side Results}\label{appB}
In this appendix, we provide the explicit form for the QCD side contributions arising from the gluon and photon EMT insertions.
\subsection{First class of gluonic contributions}\label{first class}
The explicit expression for $\langle 0 |\mathcal{T}[J_N(y)\bar{J}_N(x)]| 0 \rangle$ is given as
\allowdisplaybreaks
\begin{align}
\langle 0 |\mathcal{T}[J_N(y)\bar{J}_N(x)]| 0 \rangle&=4\ \Bigg(-\Big[\gamma_{5}\ S_u^{ca'}(y-x)\ S_d^{'\ bb'}(y-x)\ S_u^{ac'}(y-x)\ \gamma_{5}\Big]
+\Big[\gamma_5\ S_u^{cc'}(y-x)\ \gamma_5\Big]\ \text{Tr}\Big[S_u^{'\ aa'}(y-x)\ S_d^{bb'}(y-x)\Big]
\nonumber\\[7pt]&~~~~~~~
-\beta\Big[\gamma_5\ S_u^{cc'}(y-x)\Big]\ \text{Tr}\Big[\gamma_5\ S_u^{'\ aa'}(y-x)\ S_d^{bb'}(y-x)\Big]
+\beta\Big[\gamma_5\ S_u^{ca'}(y-x)\ \gamma_5\ S_d'^{bb'}(y-x)\ S_u^{ac'}(y-x)\Big]
\nonumber\\[7pt]&~~~~~~~
-\beta\ \Big[S_u^{ca'}(y-x)\ S'^{bb'}_d(y-x)\ \gamma_5\ S_u^{ac'}(y-x)\ \gamma_5\Big]
+\beta\ \Big[S^{cc'}_u(y-x)\ \gamma_5\Big]\ \text{Tr}\Big[S'^{aa'}_u(y-x)\ \gamma_5\ S^{bb'}_d(y-x)\Big]
\nonumber\\[7pt]&~~~~~~~
+\beta^2\ \Big[S_u^{ca'}(y-x)\ \gamma_5\ S'^{bb'}_d(y-x)\ \gamma_5\ S_u^{ac'}(y-x)\Big]
 -\beta^2\ \Big[S_u^{cc'}(y-x)\Big]\ \text{Tr}\Big[S'^{aa'}_u(y-x)\ \gamma_5\ S_d^{bb'}(y-x)\ \gamma_{5}\Big]\Bigg).
\label{Pi1}
\end{align}	
where $S_\psi^{ab}(x)$, with $\psi=u,d$, denotes the light-quark propagator in coordinate space, while
$S_\psi^{\prime\ ab}(x)=C\left[S_\psi^{ab}(x)\right]^T C$
represents the corresponding charge-conjugated propagator. The light-quark propagator, including perturbative and nonperturbative contributions, is expressed in coordinate space as~\cite{Agaev:2020zad},
\allowdisplaybreaks
\begin{align}
S_\psi^{ab}(x) &= i \delta^{ab} \frac{\slashed{x}}{2\pi^2 x^4} - \delta^{ab} \frac{m_\psi}{4 \pi^2 x^2} 
-\delta^{ab} \frac{\langle \bar{\psi} \psi \rangle}{12}\bigg(1-\frac{m_\psi \slashed{x}}{4}\bigg)
- \delta^{ab} \frac{x^2}{192}\left\langle  \bar{\psi} g_s \sigma G \psi \right\rangle\bigg(1-i\frac{m_\psi \slashed{x}}{6}\bigg)
\nonumber\\&
- i \frac{g_s G^{ab}_{\alpha \beta}}{32 \pi^2 x^2} \left[ \slashed{x} \sigma^{\alpha \beta} + \sigma^{\alpha \beta} \slashed{x} \right]
- i \delta^{ab} \frac{x^2 \slashed{x} g_s^2 \langle \bar{\psi} \psi \rangle^2}{7776} + \cdots,
	\label{propagator}
\end{align}
where $m_\psi$ denotes the light-quark mass. The quantities
$\langle \bar{\psi}\psi\rangle$, $\langle G^2\rangle$, and
$\langle \bar{\psi}g_s\sigma G\psi\rangle$ correspond to the quark, gluon, and mixed quark-gluon condensates, respectively.
The gluon field strength tensor appearing in the quark propagator is defined as~\cite{Asmaee:2025elo, Asmaee:2026wrk},
\allowdisplaybreaks
\begin{align}
	G_{\alpha\beta}^{ab}=G_{\alpha\beta}^{A}T_A^{ab},\qquad
	T_A=\frac{1}{2}\lambda_A,\qquad
	G^2=G_{\alpha\beta}^{A}G_{\alpha\beta}^{A},
\end{align}
where $a,b=1,2,3$ and $A=1,\dots,8$ denote the color indices of the quark and gluon fields, respectively. Here, $\lambda_A$ are the Gell-Mann matrices and $T_A$ are the generators of the $SU(3)$ color group in the fundamental representation.
The vacuum expectation value of two gluon field strength tensors in the adjoint color representation is given by~\cite{Asmaee:2025elo, Asmaee:2026wrk},
\allowdisplaybreaks
\begin{align}
\langle G_{\alpha^\prime\beta^\prime}^{A_1} G_{\alpha\beta}^{A_2} \rangle= \frac{\delta^{A_1 A_2}}{96} \ \langle G^2 \rangle \left( g_{\alpha^\prime\alpha} g_{\beta^\prime\beta} - g_{\alpha^\prime\beta} g_{\beta^\prime\alpha} \right).
\label{condensation gluon}
\end{align}
\subsection{Second class of gluonic contributions}\label{second class}
The resulting expression  $\Pi_{\mu\nu}^{g}(x,y)$ for the second class contribution, obtained from the gluon-induced correction to the light-quark propagators, is given as
\allowdisplaybreaks
\begin{align}
	\Pi_{\mu\nu}^{g}(x,y)
	=4&\, \Bigg(
	-\Big[\gamma_{5}\, \Delta S_u^{ca',\, g}(y-x)\, S_d^{'\, bb'}(y-x)\, S_u^{ac'}(y-x)\, \gamma_{5}\Big]
	-\Big[\gamma_{5}\, S_u^{ca'}(y-x)\, \Delta S_d^{'\, bb',\, g}(y-x)\, S_u^{ac'}(y-x)\, \gamma_{5}\Big]
	\nonumber\\[7pt]&~~~
	-\Big[\gamma_{5}\, S_u^{ca'}(y-x)\, S_d^{'\, bb'}(y-x)\, \Delta S_u^{ac',\, g}(y-x)\, \gamma_{5}\Big]
	+\Big[\gamma_5\, \Delta S_u^{cc',\, g}(y-x)\, \gamma_5\Big]\, \text{Tr}\Big[S_u^{'\, aa'}(y-x)\, S_d^{bb'}(y-x)\Big]
	\nonumber\\[7pt]&~~~
	+\Big[\gamma_5\, S_u^{cc'}(y-x)\, \gamma_5\Big]\, \text{Tr}\Big[\Delta S_u^{'\, aa',\, g}(y-x)\, S_d^{bb'}(y-x)\Big]
	+\Big[\gamma_5\, S_u^{cc'}(y-x)\, \gamma_5\Big]\, \text{Tr}\Big[S_u^{'\, aa'}(y-x)\, \Delta S_d^{bb',\, g}(y-x)\Big]
	\nonumber\\[7pt]&~~~
	-\beta\Big[\gamma_5\, \Delta S_u^{cc',\, g}(y-x)\Big]\, \text{Tr}\Big[\gamma_5\, S_u^{'\, aa'}(y-x)\, S_d^{bb'}(y-x)\Big]
	-\beta\Big[\gamma_5\, S_u^{cc'}(y-x)\Big]\, \text{Tr}\Big[\gamma_5\, \Delta S_u^{'\, aa',\, g}(y-x)\, S_d^{bb'}(y-x)\Big]
	\nonumber\\[7pt]&~~~
	-\beta\Big[\gamma_5\, S_u^{cc'}(y-x)\Big]\, \text{Tr}\Big[\gamma_5\, S_u^{'\, aa'}(y-x)\, \Delta S_d^{bb',\, g}(y-x)\Big]
	+\beta\Big[\gamma_5\, \Delta S_u^{ca',\, g}(y-x)\, \gamma_5\, S_d'^{bb'}(y-x)\, S_u^{ac'}(y-x)\Big]
	\nonumber\\[7pt]&~~~
	+\beta\Big[\gamma_5\, S_u^{ca'}(y-x)\, \gamma_5\, \Delta S_d'^{bb',\, g}(y-x)\, S_u^{ac'}(y-x)\Big]
	+\beta\Big[\gamma_5\, S_u^{ca'}(y-x)\, \gamma_5\, S_d'^{bb'}(y-x)\, \Delta S_u^{ac',\, g}(y-x)\Big]
	\nonumber\\[7pt]&~~~
	-\beta\,\Big[\Delta S_u^{ca',\, g}(y-x)\, S'^{bb'}_d(y-x)\, \gamma_5\, S_u^{ac'}(y-x)\, \gamma_5\Big]
	-\beta\,\Big[S_u^{ca'}(y-x)\, \Delta S'^{bb',\, g}_d(y-x)\, \gamma_5\, S_u^{ac'}(y-x)\, \gamma_5\Big]
	\nonumber\\[7pt]&~~~
	-\beta\,\Big[S_u^{ca'}(y-x)\, S'^{bb'}_d(y-x)\, \gamma_5\, \Delta S_u^{ac',\, g}(y-x)\, \gamma_5\Big]
	+\beta\,\Big[\Delta S^{cc',\, g}_u(y-x)\, \gamma_5\Big]\, \text{Tr}\Big[S'^{aa'}_u(y-x)\, \gamma_5\, S^{bb'}_d(y-x)\Big]
	\nonumber\\[7pt]&~~~
	+\beta\,\Big[ S^{cc'}_u(y-x)\, \gamma_5\Big]\, \text{Tr}\Big[\Delta S'^{aa',\, g}_u(y-x)\, \gamma_5\, S^{bb'}_d(y-x)\Big]
	+\beta\,\Big[S^{cc'}_u(y-x)\, \gamma_5\Big]\, \text{Tr}\Big[S'^{aa'}_u(y-x)\, \gamma_5\, \Delta S^{bb',\, g}_d(y-x)\Big]
	\nonumber\\[7pt]&~~~
	+
	\beta^2\,\Big[ \Delta S_u^{ca',\, g}(y-x)\, \gamma_5\, S'^{bb'}_d(y-x)\, \gamma_5\, S_u^{ac'}(y-x)\Big]
	+
	\beta^2\,\Big[S_u^{ca'}(y-x)\, \gamma_5\,  \Delta S'^{bb',\, g}_d(y-x)\, \gamma_5\, S_u^{ac'}(y-x)\Big]
	\nonumber\\[7pt]&~~~
	+
	\beta^2\,\Big[S_u^{ca'}(y-x)\, \gamma_5\, S'^{bb'}_d(y-x)\, \gamma_5\,  \Delta S_u^{ac',\, g}(y-x)\Big]
	-\beta^2\,\Big[\Delta S_u^{cc',\, g}(y-x)\Big]\, \text{Tr}\Big[S'^{aa'}_u(y-x)\, \gamma_5\, S_d^{bb'}(y-x)\, \gamma_{5}\Big]
	\nonumber\\[7pt]&~~~
	-\beta^2\,\Big[S_u^{cc'}(y-x)\Big]\, \text{Tr}\Big[\Delta S'^{aa',\, g}_u(y-x)\, \gamma_5\, S_d^{bb'}(y-x)\, \gamma_{5}\Big]
	-\beta^2\,\Big[S_u^{cc'}(y-x)\Big]\, \text{Tr}\Big[S'^{aa'}_u(y-x)\, \gamma_5\, \Delta S_d^{bb',\, g}(y-x)\, \gamma_{5}\Big]\Bigg).
	\label{Pi2}
\end{align}		
\subsection{Photonic contributions}\label{contribution photon}
After performing the Fourier transformation of the coordinate-space light-quark propagators and employing the momentum-space representation of the free photon propagator, the photon-induced correction to the light-quark propagator can be written as
\allowdisplaybreaks
\begin{align}
\Delta S_{u}^{ca',\,\gamma}&=-\frac{e_u^2}{2}\int d^4z_1\int d^4z_2\left[-\int\frac{d^4k_1}{(2\pi)^4}\frac{d^4k_2}{(2\pi)^4}\frac{N_{\alpha\beta\mu\nu}(k_1,k_2)}
{(k_1^2+i\epsilon)(k_2^2+i\epsilon)}e^{-ik_1\cdot z_1}e^{-ik_2\cdot z_2}\right]
\nonumber\\[4pt]&\qquad\times
\Bigg[\int\frac{d^4p_1}{(2\pi)^4}e^{-ip_1\cdot(y-z_1)}S_u^{ci}(p_1)\gamma^\alpha\int\frac{d^4p_2}{(2\pi)^4}e^{-ip_2\cdot(z_1-z_2)}S_u^{ij}(p_2)\gamma^\beta\int\frac{d^4p_3}{(2\pi)^4}e^{-ip_3\cdot(z_2-x)}S_u^{ja'}(p_3)
\nonumber\\&\qquad~~~~~
+\int\frac{d^4p_1}{(2\pi)^4}e^{-ip_1\cdot(y-z_2)}S_u^{ci}(p_1)\gamma^\beta\int\frac{d^4p_2}{(2\pi)^4}
e^{-ip_2\cdot(z_2-z_1)}S_u^{ij}(p_2)\gamma^\alpha\int\frac{d^4p_3}{(2\pi)^4}e^{-ip_3\cdot(z_1-x)}S_u^{ja'}(p_3)\Bigg],
\label{Su1-photon-momentum}
\end{align}
where the momentum-dependent tensor $N_{\alpha\beta\mu\nu}(k_1,k_2)$ is defined as
\allowdisplaybreaks
\begin{align}
N_{\alpha\beta\mu\nu}(k_1,k_2)&=\Big[(k_{1\mu}g_{\alpha\rho}-k_{1\rho}g_{\alpha\mu})(k_{2\nu}g_{\beta}^{\ \rho}-k_2^{\rho}g_{\beta\nu})+(k_{1\nu}g_{\alpha\rho}-k_{1\rho}g_{\alpha\nu})(k_{2\mu}g_{\beta}^{\ \rho}-k_2^{\rho}g_{\beta\mu})\Big]
\nonumber\\[7pt]&~~~
-\frac14 g_{\mu\nu}\Big[(k_{1\rho}g_{\alpha\sigma}-k_{1\sigma}g_{\alpha\rho})(k_2^{\rho}g_{\beta}^{\ \sigma}-k_2^{\sigma}g_{\beta}^{\ \rho})+(k_{2\rho}g_{\alpha\sigma}-k_{2\sigma}g_{\alpha\rho})
(k_1^{\rho}g_{\beta}^{\ \sigma}-k_1^{\sigma}g_{\beta}^{\ \rho})\Big].
\end{align}	
The explicit expression for $\Pi_{\mu\nu}^{\gamma}$ appearing in Eq.~\eqref{Pi-QCD side-photon} is given by	
\allowdisplaybreaks
\begin{align}
	\Pi_{\mu\nu}^{\gamma}
	=4&\, \Bigg(
	-\Big[\gamma_{5}\, \Delta S_u^{ca',\, \gamma}\, S_d^{'\, bb'}\, S_u^{ac'}\, \gamma_{5}\Big]
	-\Big[\gamma_{5}\, S_u^{ca'}\, \Delta S_d^{'\, bb',\, \gamma}\, S_u^{ac'}\, \gamma_{5}\Big]
	-\Big[\gamma_{5}\, S_u^{ca'}\, S_d^{'\, bb'}\, \Delta S_u^{ac',\, \gamma}\, \gamma_{5}\Big]
	\nonumber\\[7pt]&~~~
	+\Big[\gamma_5\, \Delta S_u^{cc',\, \gamma}\, \gamma_5\Big]\, \text{Tr}\Big[S_u^{'\, aa'}\, S_d^{bb'}\Big]
	+\Big[\gamma_5\, S_u^{cc'}\, \gamma_5\Big]\, \text{Tr}\Big[\Delta S_u^{'\, aa',\, \gamma}\, S_d^{bb'}\Big]
	+\Big[\gamma_5\, S_u^{cc'}\, \gamma_5\Big]\, \text{Tr}\Big[S_u^{'\, aa'}\, \Delta S_d^{bb',\, \gamma}\Big]
	\nonumber\\[7pt]&~~~
	-\beta\Big[\gamma_5\, \Delta S_u^{cc',\, \gamma}\Big]\, \text{Tr}\Big[\gamma_5\, S_u^{'\, aa'}\, S_d^{bb'}\Big]
	-\beta\Big[\gamma_5\, S_u^{cc'}\Big]\, \text{Tr}\Big[\gamma_5\, \Delta S_u^{'\, aa',\, \gamma}\, S_d^{bb'}\Big]
	-\beta\Big[\gamma_5\, S_u^{cc'}\Big]\, \text{Tr}\Big[\gamma_5\, S_u^{'\, aa'}\, \Delta S_d^{bb',\, \gamma}\Big]
	\nonumber\\[7pt]&~~~
	+\beta\Big[\gamma_5\, \Delta S_u^{ca',\, \gamma}\, \gamma_5\, S_d'^{bb'}\, S_u^{ac'}\Big]
	+\beta\Big[\gamma_5\, S_u^{ca'}\, \gamma_5\, \Delta S_d'^{bb',\, \gamma}\, S_u^{ac'}\Big]
	+\beta\Big[\gamma_5\, S_u^{ca'}\, \gamma_5\, S_d'^{bb'}\, \Delta S_u^{ac',\, \gamma}\Big]
	\nonumber\\[7pt]&~~~
	-\beta\,\Big[\Delta S_u^{ca',\, \gamma}\, S'^{bb'}_d\, \gamma_5\, S_u^{ac'}\, \gamma_5\Big]
	-\beta\,\Big[S_u^{ca'}\, \Delta S'^{bb',\, \gamma}_d\, \gamma_5\, S_u^{ac'}\, \gamma_5\Big]
	-\beta\,\Big[S_u^{ca'}\, S'^{bb'}_d\, \gamma_5\, \Delta S_u^{ac',\, \gamma}\, \gamma_5\Big]
	\nonumber\\[7pt]&~~~
	+\beta\,\Big[\Delta S^{cc',\, \gamma}_u\, \gamma_5\Big]\, \text{Tr}\Big[S'^{aa'}_u\, \gamma_5\, S^{bb'}_d\Big]
	+\beta\,\Big[ S^{cc'}_u\, \gamma_5\Big]\, \text{Tr}\Big[\Delta S'^{aa',\, \gamma}_u\, \gamma_5\, S^{bb'}_d\Big]
	+\beta\,\Big[S^{cc'}_u\, \gamma_5\Big]\, \text{Tr}\Big[S'^{aa'}_u\, \gamma_5\, \Delta S^{bb',\, \gamma}_d\Big]
	\nonumber\\[7pt]&~~~
	+
	\beta^2\,\Big[ \Delta S_u^{ca',\, \gamma}\, \gamma_5\, S'^{bb'}_d\, \gamma_5\, S_u^{ac'}\Big]
	+
	\beta^2\,\Big[S_u^{ca'}\, \gamma_5\,  \Delta S'^{bb',\, \gamma}_d\, \gamma_5\, S_u^{ac'}\Big]
	+
	\beta^2\,\Big[S_u^{ca'}\, \gamma_5\, S'^{bb'}_d\, \gamma_5\,  \Delta S_u^{ac',\, \gamma}\Big]
	\nonumber\\[7pt]&~~~
	-\beta^2\,\Big[\Delta S_u^{cc',\, \gamma}\Big]\, \text{Tr}\Big[S'^{aa'}_u\, \gamma_5\, S_d^{bb'}\, \gamma_{5}\Big]
	-\beta^2\,\Big[S_u^{cc'}\Big]\, \text{Tr}\Big[\Delta S'^{aa',\, \gamma}_u\, \gamma_5\, S_d^{bb'}\, \gamma_{5}\Big]
	-\beta^2\,\Big[S_u^{cc'}\Big]\, \text{Tr}\Big[S'^{aa'}_u\, \gamma_5\, \Delta S_d^{bb',\, \gamma}\, \gamma_{5}\Big]\Bigg).
	\label{Pigamma}
\end{align}					
\section{Derivation of the Mass and Scalar Radii}\label{app:radii}
In this appendix, Eq.~\eqref{radii-j} for the mass and scalar radii is derived. The usual equations in the literature \cite{Guo:2023pqw,Ji:2021mtz} are obtained from the complete proton EMT. Without the quark contribution, the GFF $\bar{c}_j(t)$ must be included.
\subsection{Derivation of the Mass Radius}
For each EMT contribution $j$, the mass radius is defined from the second moment of its energy density,
\allowdisplaybreaks
\begin{align}
\langle r_m^2\rangle_j=\frac{\displaystyle\int d^3r r^2\varepsilon_j(r)}{\displaystyle\int d^3r\ \varepsilon_j(r)},
\label{MassRadiusDef}
\end{align}
where $\varepsilon_j(r)$ is the energy density, which is defined in the Breit frame from the temporal component of the EMT as
\allowdisplaybreaks
\begin{align}
\varepsilon_j(r)=T^{00}_{j}(r).
\label{density}
\end{align}
The three-dimensional Fourier transform of a GFF is defined as
\allowdisplaybreaks
\begin{align}
[f(t)]_{\mathrm{FT}}=\int\frac{d^3\boldsymbol{\Delta}}{(2\pi)^3}e^{-i\boldsymbol{\Delta}\cdot\mathrm{r}}f(t), \qquad t=\Delta^2=-\boldsymbol{\Delta}^{2}.
	\label{FT}
\end{align}
Using the EMT matrix element in Eq.~\eqref{matrix element}, the energy density in Eq.~\eqref{density} is
\allowdisplaybreaks
\begin{align}
\varepsilon_j(r)=m\Big[A_j(t)+\bar{c}_j(t)-\frac{t}{4m^2}\left(A_j(t)-2J_j(t)+D_j(t)\right)\Big]_{\mathrm{FT}}.
\label{EnergyDensity}
\end{align}
The normalization of the energy density is obtained from the zeroth moment of the Fourier transform,
\allowdisplaybreaks
\begin{align}
\displaystyle\int d^3r\ \varepsilon_j(r)=m\big[A_j(0)+\bar{c}_j(0)\big].
\end{align}
Expanding the GFFs around the forward limit, $t=0$, 
\allowdisplaybreaks
\begin{align}
F_j(t)=F_j(0)+tF_j'(0)+{\cal{O}}(t^2),
\label{expand}
\end{align}
where $F_j=\{A_j,\bar{c}_j,J_j,D_j\}$. Using the standard property of the three-dimensional Fourier transform defined in Eq.~\eqref{FT},
\allowdisplaybreaks
\begin{align}
\displaystyle\int d^3r\ r^2[f(t)]_{\mathrm{FT}}=6\left.\frac{df(t)}{dt}\right|_{t=0},
\label{integral FT}
\end{align}
the second moment of the energy density is calculated. Substituting the resulting relation into Eq.~\eqref{MassRadiusDef},
\allowdisplaybreaks
\begin{align}
\langle r_m^2\rangle_j=\frac{1}{A_j(0)+\bar{c}_j(0)}\left[6\left.\frac{dA_j(t)}{dt}\right|_{t=0}
+6\left.\frac{d\bar{c}_j(t)}{dt}\right|_{t=0}-\frac{3}{2m^2}\left(A_j(0)-2J_j(0)+D_j(0)\right)\right].
\label{rm-final}
\end{align}
For comparison, Eq.~\eqref{rm-final} reduces to the commonly used relation for the complete proton EMT.
Using the forward limit relations from previous studies,
\begin{align}
	J_q(0)+J_g(0)=\frac{1}{2},\qquad
	A_q(0)+A_g(0)=1,\qquad
	\bar c_q(0)+\bar c_g(0)=0,
	\label{relation}
\end{align}
and substituting these relations into Eq.~\eqref{rm-final}, the result agrees with the form used in literature \cite{Guo:2023pqw,Ji:2021mtz}, with $D(0)=4C(0)$,
\allowdisplaybreaks
\begin{align}
\langle r_m^2\rangle=\frac{1}{A(0)}\left[6\left.\frac{dA(t)}{dt}\right|_{t=0}
-\frac{6}{m^2}C(0)\right].
\label{rm-final-qg}
\end{align}
\subsection{Derivation of the Scalar Radius}
The scalar radius is defined as the normalized second moment of the scalar density, as follows:
\allowdisplaybreaks
\begin{align}
\langle r_s^2\rangle_j=\frac{\displaystyle\int d^3r\ r^2\rho_j(r)}{\displaystyle\int d^3r\ \rho_j(r)},
	\label{ScalarRadiusDef}
\end{align}
where $\rho_j(r)$ is the scalar density, which is defined from the trace of the EMT. Since the EMT has two Lorentz indices, its trace is obtained by contracting it with the metric tensor,
\allowdisplaybreaks
\begin{align}
T^\mu_{\ \mu}=g^{\mu\nu}T_{\mu\nu}.
\end{align}
Using Eq.~\eqref{matrix element}, together with the on-shell conditions $p^2=p'^2=m^2$, and the Gordon identity, the trace of the EMT matrix element is obtained as
\allowdisplaybreaks
\begin{align}
	\left\langle  N(p',s')\left| g_{\mu\nu}T^{\mu\nu}_j(0)\right| N(p,s)\right\rangle =m\
	\bar{u}(p',s')&\Bigg\{ A_j(t)+4\bar{c}_j(t)+\frac{t}{2m^2}J_j(t)-\frac{t}{4m^2}A_j(t)
	-\frac{3t}{4m^2}D_j(t)\Bigg\}u(p,s).
	\label{matrix element1}
\end{align}	
The scalar density is obtained from Eq.~\eqref{matrix element1} in the Breit frame and is defined by its three-dimensional Fourier transform,
\allowdisplaybreaks
\begin{align}
\rho_j(r)=m\Big[A_j(t)+4\bar{c}_j(t)+\frac{t}{2m^2}J_j(t)-\frac{t}{4m^2}A_j(t)-\frac{3t}{4m^2}D_j(t)\Big]_{\mathrm{FT}}.
\end{align}
The normalization of the scalar density is determined by the Fourier transform at $t=0$,
\allowdisplaybreaks
\begin{align}
\displaystyle\int d^3r\ \rho_j(r)=m\Big[A_j(0)+4\bar{c}_j(0)\Big].
\label{denominator}
\end{align}
Using the GFF expansion around the forward limit in Eq.~\eqref{expand} and the Fourier transform in Eq.~\eqref{integral FT}, the second moment of the scalar density is
\begin{align}
\displaystyle\int d^3r r^2\rho_j(r)=6m\Bigg[\left.\frac{dA_j(t)}{dt}\right|_{t=0}
+4\left.\frac{d\bar{c}_j(t)}{dt}\right|_{t=0}
+\frac{J_j(0)}{2m^2}-\frac{A_j(0)}{4m^2}-\frac{3}{4m^2}D_j(0))\Bigg].
\label{numerator}
\end{align}
Substituting Eqs.~\eqref{denominator} and \eqref{numerator} into Eq.~\eqref{ScalarRadiusDef}, one obtains
\allowdisplaybreaks
\begin{align}
\langle r_s^2\rangle_j=\frac{1}{A_j(0)+4\bar{c}_j(0)}
\Bigg[6\left.\frac{dA_j(t)}{dt}\right|_{t=0}+24\left.\frac{d\bar{c}_j(t)}{dt}\right|_{t=0}
-\frac{3}{2m^2}\Big(A_j(0)-2J_j(0)+3D_j(0)\Big)\Bigg].
\label{ScalarRadiusFinal}
\end{align}
For comparison, we consider the complete proton EMT, including both quark and gluon contributions. Using the forward limit relations in Eq.~\eqref{relation} in Eq.~\eqref{ScalarRadiusFinal}, the usual expression for the scalar radius is obtained \cite{Guo:2023pqw,Ji:2021mtz}, with $D(t)=4C(t)$.
\begin{align}
	\langle r_s^2\rangle
	=
	\frac{1}{A(0)}
	\left[
	6A'(0)
	-\frac{18}{m^2}C(0)
	\right].
\end{align}

\end{document}